\documentclass[aps,pra,twocolumn,superscriptaddress]{revtex4}
\usepackage{amsmath}
\usepackage{graphicx}		
\usepackage{graphics}
\usepackage[usenames]{color}

\begin{document}

\newcommand{\diff}[2]{\frac{\partial #1}{\partial #2}}
\newcommand{\secdiff}[2]{\frac{\partial^2 #1}{\partial #2^2}}
\newcommand{\comment}[1]{{\bf \textcolor{red}{ #1}}}
\newcommand{\BDEcomment}[1]{{\bf \textcolor{blue}{ #1}}}
\newcommand{\NPMcomment}[1]{{\bf \textcolor{ForestGreen}{ #1}}}

\title{Three-Body Recombination in One Dimension}

\author{N.~P.~Mehta}
\affiliation{JILA, University of Colorado, Boulder, CO 80309-0440}
\email[]{mehtan@jilau1.colorado.edu}
\author{B.~D.~Esry}
\affiliation{Department of Physics, Kansas State University, Manhattan, KS 66506}
\email[]{esry@phys.ksu.edu}
\author{C.~H.~Greene}
\affiliation{Department of Physics and JILA, University of Colorado, Boulder, CO 80309-0440}
\email[]{chris.greene@colorado.edu}

\date{\today}

\begin{abstract}
We study the three-body problem in one dimension for both zero and finite range interactions using the 
adiabatic hyperspherical approach.  Particular emphasis is placed
on the threshold laws for recombination, which are derived for all combinations of the parity and exchange symmetries.
For bosons, we provide a numerical demonstration of several universal features that appear in the three-body system, 
and discuss how certain universal features in three dimensions are different in one dimension.  
We show that the probability for inelastic processes vanishes as the range of the pair-wise interaction 
is taken to zero and demonstrate numerically that the recombination threshold law manifests itself for large scattering length.  
\end{abstract}


\maketitle

\section{Introduction\label{intro}}

One-dimensional (1-D) few-body and many-body systems have been the subject of intense theoretical study for many 
years~\cite{mcguire,yang,lieb_liniger,thacker,girardeau,tonks}.  This is largely because certain one-dimensional 
problems admit exact solutions using the Bethe ansatz.  These theoretical studies are gaining increasing attention
due to the experimental realization of effective 1-D geometries
in tightly confined cylindrical trap geometries~\cite{gorlitz_etal,
tolra_etal,kinoshita2006qnc,kinoshita2004otg,meyrath2005bec,esteve2006odf}.  
Atoms in such traps are essentially free to propagate in one coordinate while being restricted to the lowest cylindrical mode in 
the transverse radial coordinate.  Olshanii has shown~\cite{olshanii1998asp} how the 1-D scattering length (which we call $a$)
is related to the three-dimensional $s$-wave scattering length $a_{3D}$ and to the oscillator length $a_{\perp}$ in 
the confined radial direction.  This identification allows a connection with the intensively-studied 1-D
zero range model with two-body interactions
\begin{equation}
\label{v_delta}
V(x)=g\delta(x).
\end{equation}
The coupling constant $g$ is renormalized to account for virtual 
transitions to excited radial modes which appear as closed channels 
in the unrestricted coordinate $x$.  Early analytic calculations by McGuire~\cite{mcguire} using the 
interaction Eq.~(\ref{v_delta}) gave vanishing probability for
all inelastic events such as collision induced break-up ($AA+A\rightarrow A+A+A$) and three-body 
recombination ($A+A+A\rightarrow AA+A$).  Section~\ref{subsection:zrp_vs_finite} discusses how finite-range 
interactions break the integrability of zero-range models, and also how the probability for recombination and
break-up behave with respect to the range of the interaction.  In Section~\ref{discussion}, we comment 
briefly on the relevance of our 1-D model to physical systems in actual atomic waveguides.

In three dimensions (3-D), universal features occur when the scattering length is the
largest length scale (see~\cite{braaten2006ufb} and references therein).  
Adiabatic hyperspherical studies in three dimensions have provided a great deal 
of insight into such universal behavior~\cite{nielsen1999ler,nielsen_etal,esry1999rta,suno2002tbr}.
For example, the Efimov effect in 3-D appears as a hyperradial potential curve in 
the region $r_0 \ll R \ll |a|$ that is attractive and varies as $R^{-2}$.  
Since it has a supercritical coefficient, this potential has (as $a_{3D}\rightarrow \infty$) 
an infinity of long-range bound states spaced geometrically by a universal 
constant~\cite{dincao2005mee}.  One of our goals here is to determine what kind of universal behavior, if any, appears
in one dimension.  We discuss these issues in Section~\ref{1Duniversality}.  

Finally, a major portion of this work deals with the threshold laws for recombination in 1-D.  
We outline how these
laws can be extracted from the asymptotic form of the adiabatic potential curves with a generalized
Wigner analysis~\cite{wigner1948bcs,ThresholdReview,esry2001tlt}, and show that --- as with any such 
analysis --- they are independent of
the short-range properties of the interactions.  
This work closely parallels the 3-D analysis we carried out in Ref.~\cite{esry2001tlt}.
Threshold laws are found for all 
combinations of the parity and exchange symmetries, including the cases
where only two of the three particles are identical.  In order to demonstrate the threshold behavior, we
present numerical calculations for bosons.  

\section{Hyperspherical coordinates}
\label{coordinates}

Since hyperspherical coordinates and the adiabatic hyperspherical representation 
play a central role in this paper, but may be unfamiliar to some readers, we will
briefly discuss the important points.  For more detailed information, we refer the
interested reader to Refs.~\cite{nielsen_etal,CDLReview,suno2002tbr}.

We separate the center
of mass motion from the relative motion using Jacobi coordinates,
\begin{equation}
\label{app:jacobi}
x_{12} = x_1-x_2 ~~~ {\rm and} ~~~  x_{12,3} = \frac{m_1 x_1+m_2 x_2}{m_1+m_2} - x_3.
\end{equation}
The positions $x_i$ locate each particle relative to some laboratory-fixed
origin, and $m_i$ are their masses.  The Jacobi coordinates $x_{12}$ and $x_{12,3}$
constitute a Cartesian coordinate system.  Transforming these to polar coordinates
$(R,\phi)$,
\begin{equation}
\mu R^2 = \mu_{12}x_{12}^2 + \mu_{12,3}x_{12,3}^2 ~~~ {\rm and} ~~~ \tan\phi = \sqrt{\frac{\mu_{12,3}}{\mu_{12}}}\frac{x_{12,3}}{x_{12}},
\end{equation}
gives the hyperspherical coordinate system.  The reduced masses are
\begin{equation}
\mu_{12} = \frac{m_1m_2}{m_1+m_2} ~~~ {\rm and} ~~~ \mu_{12,3}=\frac{(m_1+m_2)m_3}{m_1+m_2+m_3},
\end{equation}
and we choose the three-body reduced mass $\mu$ to be $\sqrt{\mu_{12}\mu_{12,3}}$.

With the above definitions, we can find the interparticle distances to be
\begin{equation}
\label{uneq_jacobi}
|x_i-x_j| = d_{ij}R\,|\!\sin(\phi-\phi_{ij})|,
\end{equation}
where we have introduced the constants:
\begin{align}
\label{uneq_dmassconst}
\frac{d_{12}}{\sqrt{\mu}}&=\sqrt{\frac{1}{\mu_{12}}}
\nonumber\\
\frac{d_{23}}{\sqrt{\mu}}&=\sqrt{\frac{1}{\mu_{12,3}} + \left(\frac{m_1}{m_{12}}\right)^2\frac{1}{\mu_{12}}} 
\nonumber\\
\frac{d_{31}}{\sqrt{\mu}}&=\sqrt{\frac{1}{\mu_{12,3}} + \left(\frac{m_2}{m_{12}}\right)^2\frac{1}{\mu_{12}}} .
\end{align}
The coalescence points --- where the interparticle distances are zero --- are defined as
\begin{align}
\label{coalescense_pts}
\phi_{12}&=\frac{\pi}{2}
\nonumber\\
\tan{\phi_{23}}&=\frac{m_1}{m_{12}}\sqrt{\frac{\mu_{12,3}}{\mu_{12}}} 
\nonumber\\
\tan{\phi_{31}}&=\frac{m_2}{m_{12}}\sqrt{\frac{\mu_{12,3}}{\mu_{12}}} .
\end{align}
If all three particles have the same mass, then $d_{ij}=\sqrt{2/\sqrt{3}}$ and $\phi_{23}=-\phi_{31}=\pi/6$.
If there are only two identical particles, then it is convenient to label them 1 and 2.  In this case,
$d_{23}=d_{31}$ and $\phi_{23}=-\phi_{31}$.  

Since the hyperradius $R$ is the only length scale in the system, giving the
overall size of the system, it is natural to treat it as a ``slow'' coordinate
for an adiabatic representation~\cite{macek1968pas}.  The Hamiltonian
for the system can be written as
\begin{equation}
H = -\frac{1}{2 \mu R}\frac{\partial}{\partial R}\left(R\frac{\partial}{\partial R}\right) + H_{\rm ad}(R, \phi)
\label{FullHam}
\end{equation}
where
\begin{equation}
H_{\rm ad}(R, \phi) = 
-\frac{1}{2 \mu R^2}\frac{\partial^2}{\partial\phi^2} + V(R,\phi)
\label{AdHam}
\end{equation}
and $V$ includes all of the interactions.
Atomic units have been used here and will be used throughout this work.
The adiabatic representation is then defined by the equation
\begin{equation}
H_{\rm ad}(R, \phi) \Phi_\nu(R; \phi) = U_\nu(R) \Phi_\nu(R; \phi).
\label{AdEqn}
\end{equation}
The eigenfunctions $\Phi_\nu(R; \phi)$ are the channel functions, and the
eigenvalues $U_\nu(R)$ the potential curves corresponding to each channel.

In the limit $R\rightarrow\infty$, the potentials approach either the energies of
the bound diatomic molecules for the recombination (two-body) channels, or zero energy
for the three-body continuum channels.  In this limit, the channels are uncoupled, although
there must be coupling at smaller $R$ for any inelastic transition such as
recombination to occur.  For the purposes of determining the threshold
laws, however, we need only know that the channels in this representation become uncoupled
asymptotically.

\section{Threshold Behavior}
\label{section:threshold}

At ultracold temperatures, the dominant character of a given scattering process is controlled by its 
threshold behavior, and the adiabatic hyperspherical picture readily yields this
behavior~\cite{esry2001tlt}.  We show this by first solving 
the adiabatic equation Eq.~(\ref{AdEqn}) while taking into account the
appropriate exchange symmetry and parity.  
In the limit $R\rightarrow\infty$, all of the adiabatic potentials 
take the general form $U_{\kappa}\rightarrow\kappa^2/(2\mu R^2)$, yielding
the hyperradial equation
\begin{equation}
\left[-\frac{1}{R}\diff{}{R}\left(R\diff{}{R}\right) + \frac{\kappa^2}{R^2} - k^2\right]F(R)=0
\end{equation}
where $k$ is related to the total energy $E$ ($E$=0 at the three-body breakup threshold) 
by $k^2=2\mu E$.  The final momentum in the two-body channel is related to 
the total energy by $k_{f}^2=2\mu(E+B_2)$, where
$B_2$ is the two-body binding energy.
This equation is simply Bessel's equation with general solution
\begin{equation}
\label{asymform}
F_\kappa(kR) = AJ_{\kappa}(kR) + BY_{\kappa}(kR).
\end{equation}
The coefficients $A$ and $B$ are determined by the usual procedure of matching
to short-range solutions and are related to the $S$-matrix.
It is more convenient for the present discussion, though, to consider
each $S$-matrix element for recombination at a fixed total energy $E$ in the form
\begin{equation}
S_{fi}(E) \sim \left<F_{2B}\Phi_{2B}|\hat{S}| F_{3B}\Phi_{3B}^\kappa\right>,
\end{equation}
where $2B$ labels the final atom-dimer channel
and $3B$ labels the initial three-body continuum state.
To determine the threshold behavior, 
we use the small argument form of the Bessel functions.
For recombination, the final momentum in the two-body channel
is non-zero and slowly varying
at the three-body threshold, but the initial momentum $k$ vanishes there.  Hence, it is
the initial channel that determines the energy dependence
of the transition probability, and it is the 
lowest three-body continuum channel, with $\kappa=\kappa_{\rm min}$,
that will dominate at threshold.  As a result,
in the limit $k\rightarrow 0$, the recombination probability must scale like
\begin{equation}
|S_{fi}|^2 \propto k^{2\kappa_{\rm min}}.
\label{S_scaling}
\end{equation}
With this scaling, we can, of course, determine the scaling behavior of
the recombination rate.  This connection will be discussed in Sec.~\ref{RateDef}.
Moreover, when the scattering length is the largest length scale, simple dimensional arguments 
(namely that a probability must be unitless), imply that
\begin{equation}
|S_{fi}|^2 \propto (ka)^{2\kappa_{\rm min}}.
\label{S_a_scaling}
\end{equation}


\subsection{Three identical bosons with $\delta$-function interactions}
\label{thresh:3ident_delta}

In this section, we will use $\delta$-function pair potentials, Eq.~(\ref{v_delta}),
to find $\kappa_{\rm min}$ for three identical bosons.  Even though the recombination
rate for such interactions actually vanishes~\cite{mcguire}, the analytic solutions of
the adiabatic equation possible with these interactions~\cite{gibson_etal,mehta1} nevertheless give $\kappa_{\rm min}$ for general
interactions.  With $\delta$-function interactions, we can treat only interacting
bosons.  The case of interacting fermions will be considered in Sec.~\ref{SqWell}.

For three identical particles, the coalescence points (e.g. $x_i-x_j=0$) form radial lines that are equally spaced by $\pi/3$ 
in the two-dimensional space spanned by the Jacobi coordinates defined in 
Eq.~(\ref{app:jacobi}).  These lines divide the coordinate space into 
six regions, each of which corresponds to a unique ordering of the three particles along the real line.  
Symmetry thus permits us to solve the adiabatic equation (\ref{AdEqn}) in just 
the region $0\le\phi\le\pi/6$ with appropriate boundary conditions (see App.~\ref{app:symmops}).

The $\delta$-function coupling constant $g$ is related to the 1-D two-body scattering 
length by $a=-1/(\mu_{2B}g)$ which is defined from the 1-D effective range
expansion in the even parity ``partial wave'',
$k\tan(\delta) = \frac{1}{a} + \frac{r_0}{2}k^2 + \ldots$~\cite{felline2003leo,olshanii1998asp}.
We require $g < 0$ so that the potential supports a two-body
bound state, and write the general solution for this channel as
\begin{equation}
\Phi(R;\phi)=A\sinh{q\phi} + B\cosh{q\phi}
\end{equation}
where $q$ is related to the potential energy through:
\begin{equation}
U(R)=-\frac{q^2}{2\mu R^2}.
\end{equation}
For even parity, we impose boundary conditions $\Phi'(0)=0$ and
\begin{equation}
\label{bc_delta}
\lim_{\epsilon\rightarrow 0^+}{\Phi'(\phi_0+\epsilon)-\Phi'(\phi_0-\epsilon)}=2\mu g\alpha R\Phi(\phi_0),
\end{equation}
where $\alpha=1/d_{12}=3^{1/4}2^{-1/2}$ and $\phi_0=\pi/6$.  We require the additional symmetry condition 
that $\Phi'(\pi/6+\epsilon) = -\Phi'(\pi/6-\epsilon)$.  These conditions lead to a transcendental
equation for $q$:
\begin{equation}
\label{delta_tran_2b}
q\tanh{\frac{q\pi}{6}}=-\mu g\alpha R .
\end{equation}
A similar analysis for the continuum solutions begins with the general solution
\begin{equation}
\Phi(R;\phi)=A\sin(\kappa\phi) + B\cos(\kappa\phi)
\end{equation}
where $\kappa$ is related to the potential energy through
\begin{equation}
U(R)=\frac{\kappa^2}{2\mu R^2}
\label{kappa_to_U}
\end{equation}
and is the solution to the following transcendental equation:
\begin{equation}
\label{ebdelta}
\kappa\tan{\frac{\kappa\pi}{6}}=\mu g \alpha R.
\end{equation}
As $R\rightarrow \infty$, the allowed solutions are $\kappa=3,9,15...$.
Note that Eq.~(\ref{delta_tran_2b}) is, of course, the analytic continuation of 
Eq.~(\ref{ebdelta}) with $\kappa \rightarrow iq$.

For odd parity solutions, the only difference is that the boundary condition at $\phi=0$ changes to $\Phi(0)=0$,
immediately giving $B=0$ and leading to 
\begin{equation}
\label{delta_tran_2b_odd}
q\coth{\frac{q\pi}{6}}=-\mu g\alpha R 
\end{equation}
and
\begin{equation}
\label{obdelta}
\kappa\cot{\frac{\kappa\pi}{6}}=\mu g \alpha R.
\end{equation}
From Eq.~(\ref{obdelta}), the allowed values $\kappa$ for odd parity bosons asymptotically are $\kappa=6,12,18 \ldots$
Note that the even and odd parity potential curves for the two-body channel become 
degenerate in the limit $R\rightarrow \infty$ as one would expect.

\subsection{Three identical particles with square-well interactions}
\label{SqWell}

In order to determine whether the $\delta$-function results of Sec.~\ref{thresh:3ident_delta} 
are general, we now consider square-well pair-wise interactions:
\begin{equation}
V(x_{ij})=
\begin{cases}
-V_0,\; &\text{if $0\le |x_{ij}| \le L$;}\\
0, &\text{otherwise}.
\end{cases} 
\end{equation}
Like the $\delta$-function interactions, the adiabatic equation remains 
analytically solvable for square-well interactions.  In addition, square-well
interactions are a good model for short --- but non-zero --- range interactions.
At the end of this section, we will point out that the results are, in fact, 
general for all short-range interactions.  Unlike the $\delta$-function interactions,
though, indistinguishable fermions can interact via square-well interactions.  
Consequently, we will be able to find $\kappa_{\rm min}$ for three identical fermions.

We thus focus on solutions that are either completely symmetric or completely antisymmetric ---
assuming as in Sec.~\ref{thresh:3ident_delta} that the particles are spin polarized such that the spatial wave function carries
all of the permutation symmetry.
As in Sec.~\ref{thresh:3ident_delta}, symmetry permits us to
to solve the adiabatic equation (\ref{AdEqn}) in a wedge of width $\frac{\pi}{6}$.  To simplify the imposition 
of boundary conditions, though, we choose the interval $\frac{\pi}{2}\le\phi\le \frac{2\pi}{3}$ so that the edge $\phi_b$
of the square-well centered at $\phi=\frac{\pi}{2}$ is simply
\begin{equation}
\cos{\phi_b}=-\left(\frac{3}{4}\right)^{\frac{1}{4}}\frac{L}{R}.
\end{equation}
This condition is invalid at small $R$ where the two-body interactions overlap, but is sufficient for 
determining the allowed $\kappa$ in the region $R\gg L$. 
Again using $U(R)=\kappa^2/(2\mu R^2)$ (\ref{kappa_to_U}) and defining $\beta^2=2\mu R^2 V_0$, the adiabatic equation becomes:
\begin{align}
\left(\secdiff{}{\phi}+\beta^2\right)\Phi &= -\kappa^2\Phi \;\; &\frac{\pi}{2}\le\phi\le\phi_b\nonumber\\
\secdiff{}{\phi}\Phi&=-\kappa^2\Phi \;\; &\phi_b\le\phi\le \frac{2\pi}{3}
\end{align}
We write the continuum solutions as
\begin{equation*}
\Phi=\!
\begin{cases}
A\sin\!{\left[\sqrt{\kappa^2+\beta^2}(\phi-\frac{\pi}{2})\right]} + &\\
    ~~~~~B\cos\!{\left[\sqrt{\kappa^2+\beta^2}(\phi-\frac{\pi}{2})\right]}, 
	&\frac{\pi}{2}\le\phi\le\phi_b \\
C\sin\!{\left[\kappa(\phi\!-\!\frac{2\pi}{3})\right]} \!+\! D\cos\!{\left[\kappa(\phi\!-\!\frac{2\pi}{3})\right]}, &\phi_b\le\phi\le\frac{2\pi}{3}
\end{cases}
\end{equation*}
A similar expression may be written for the two-body channels, but 
we want to focus on the recombination threshold behavior and thus need only 
the asymptotic behavior of the three-body channels.

For brevity, we sketch the derivation only for even-parity bosons and summarize the results for
all other symmetries in App.~\ref{app:Summary}.  For this symmetry, we impose boundary conditions:
\begin{equation}
\Phi'(\frac{\pi}{2})=\Phi'(\frac{2\pi}{3})=0
\end{equation}
which gives $A=C=0$.  Matching the log-derivatives of the wave function at $\phi=\phi_b$ yields the quantization
condition:
\begin{equation}
\sqrt{\kappa^2+\beta^2}\tan\!{\left[\sqrt{\kappa^2+\beta^2}(\phi_b\!-\!\frac{\pi}{2})\right]}=
\kappa\tan\!{\left[\kappa(\phi_b\!-\!\frac{2\pi}{3})\right]}.
\label{EvenBosonSqWell}
\end{equation}
In the limit $R\rightarrow \infty$, $\phi_b\rightarrow \frac{\pi}{2}$, 
$\sqrt{\kappa^2+\beta^2} \rightarrow \beta$, and 
\begin{equation}
\sqrt{\kappa^2+\beta^2}(\phi_b-\frac{\pi}{2})\rightarrow \sqrt{2\mu V_0}\left(\frac{3}{4}\right)^{\frac{1}{4}}L\equiv\lambda.
\end{equation}
So, in this limit, Eq.~(\ref{EvenBosonSqWell}) becomes
\begin{equation}
\label{ebsqr}
\beta\tan{\lambda}=
-\kappa\tan{\frac{\kappa\pi}{6}}
\end{equation}
Since $\beta\propto R$, this equation has the same form as Eq.~(\ref{ebdelta}), with the same consequences.  
In particular, the allowed $\kappa$ are $3,9,15...$. 


This analysis can be easily extended to find all allowed values of $\kappa$ for all symmetries (see App.~\ref{app:Summary}).
The threshold law, however, depends only on $\kappa_{\rm min}$.  Collecting this quantity for each 
symmetry, we find for bosons that
\begin{equation*}
\kappa_{\rm min}=
\begin{cases}
 3 &\text{Even Parity}\\
 6 &\text{Odd Parity} \\
\end{cases}.
\end{equation*}
in agreement with the results from Sec.~\ref{thresh:3ident_delta} for $\delta$-function interactions.
For fermions, we find
\begin{equation}
\kappa_{\rm min}=
\begin{cases}
 6 &\text{Even Parity}\\
 3 &\text{Odd Parity} \\
\end{cases}.
\end{equation}
These fermion $\kappa_{\rm min}$ are actually the same as one would find by 
symmetrizing the free particle solutions [see Eq.~(\ref{FreeIdentFermions})].
The interacting boson $\kappa_{\rm min}$ above, though, are not the same as the symmetrized
free particle solutions from Eq.~(\ref{FreeIdentBosons}).
From Eq.~(\ref{S_scaling}), we see that recombination at threshold will be dominated
by the even parity symmetry for bosons and by odd parity for fermions.  Moreover, recombination
of bosons and fermions~\cite{EsryICAP2002} in 1-D share the same threshold law.

The above analysis, in fact, generalizes to short-range potentials of any form.  That is, 
if we take the log-derivative $b$ [see Eq.~(\ref{LogDeriv})] from the two-body equation
outside the range of the interaction, then the matching condition is
\begin{equation}
\left(\frac{4}{3}\right)^\frac{1}{4} b R =
\kappa\tan{\left[\kappa(\phi_b-\frac{2\pi}{3})\right]}
\label{GenInt}
\end{equation}
for even-parity bosons.
In the limit $R\rightarrow \infty$, $b$ may be regarded as a constant so that this equation 
reduces to the same form as Eq.~(\ref{ebdelta}) or Eq.~(\ref{ebsqr}) and carries the same consequences.

\subsection{Two identical particles 
with $\delta$-function interactions}
\label{Two_delta}

For this case, all combinations of parity and exchange are considered 
assuming the spins of each particle are fixed by spin polarization.
Following the discussion in Sec.~\ref{coordinates}, we label the
two indistinguishable particles 1 and 2.
The interaction then takes the form
\begin{equation}
\label{uneq_vdelta}
V = g_S\delta(x_{12}) + g_D\left[\delta(x_{31})+\delta(x_{23})\right]
\end{equation}
where $g_S$ denotes the same-particle coupling, $g_D$ denotes the different-particle coupling,
and $x_{ij}$ are the interparticle distances.  The 
constant $\alpha$ in Eq.~(\ref{bc_delta}) is equal to $1/d_{ij}$ for each pair of particles $ij$.  
We shall use $\alpha_S$ for particles 1 and 2
and $\alpha_D$ otherwise.

By symmetry, we need only solve the adiabatic equation in the range $0\le \phi \le \pi/2$.
The $\delta$-function boundary condition Eq.~(\ref{bc_delta}) is imposed at $\phi_0=\phi_{23}$ for the
distinguishable pair [see Eq.~(\ref{coalescense_pts})] and at
$\phi_0=\pi/2$ for the indistinguishable pair.  The general
continuum solution is now conveniently written as
\begin{equation}
\Phi=\!
\begin{cases}	
A\cos{(\kappa\phi)}+B\sin{(\kappa\phi)}, &0\le\phi\le\phi_{23};\\
C\cos{[\kappa(\phi\!-\!\frac{\pi}{2})]}+D\sin{[\kappa(\phi\!-\!\frac{\pi}{2})]}, &\phi_{23}\le\phi\le\frac{\pi}{2}.
\end{cases}
\end{equation}
We are now prepared to determine the asymptotically allowed values of $\kappa$ for specific symmetries.


For brevity, we outline the derivation only for even-parity bosons and 
summarize the results for all cases in App.~\ref{app:Summary}.  
As before, even-parity bosons require the boundary conditions
\begin{equation}
\Phi'(0)=0 \qquad \text{and} \qquad \Phi'(\frac{\pi}{2}+\epsilon)=-\Phi'(\frac{\pi}{2}-\epsilon).
\end{equation}
These conditions immediately give $B=0$ and yield the quantization condition:
\begin{widetext}
\begin{equation}
\label{uneq_tran_eb}
\left(-\kappa + \frac{2}{\kappa}\mu^2 g^2\alpha_S\alpha_D R^2 \right)\tan\!\left[\kappa(\phi_{23}-\frac{\pi}{2})\right] - \mu g\alpha_S R\tan{\kappa\phi_{23}}
	\tan\!\left[\kappa(\phi_{23}-\frac{\pi}{2})\right] + \kappa\tan{\kappa\phi_{23}} = \mu g(\alpha_S + 2\alpha_D)R.
\end{equation}
\end{widetext}
In the limit $R\rightarrow\infty$, the $R^2$ term dominates, and this equation reduces to
\begin{equation}
\frac{1}{\kappa}\tan{\kappa(\phi_{23}-\frac{\pi}{2})}=0,
\end{equation}
yielding solutions $\kappa=n\pi/|\phi_{23}-\frac{\pi}{2}|$ for $n=1,2,3...$.  Careful inspection of Eq.~(\ref{uneq_tran_eb})
leads to the additional solution
\begin{equation}
\kappa\tan{\kappa\phi_{23}}\rightarrow\infty.
\end{equation}
This equation has solutions $\kappa=(n+1/2)\pi/|\phi_{23}|$ for $n=0,1,2...$.  

Similar analyses can be carried out for
all combinations of parity and exchange (see App.~\ref{app:Summary}).
The resulting $\kappa_{\rm min}$ for each symmetry are summarized below.
As discussed above, $\kappa_{min}$ 
determines the threshold law for recombination and is, for two identical particles in general, 
an irrational number that depends only on the masses.
Moreover, since $\kappa_{min}$ is always a nonzero quantity, we 
find that the recombination rate is never a constant at threshold in 1-D.

For bosons we have
\begin{equation*}
\kappa_{\rm min}=
\begin{cases}
\begin{cases}
\frac{\pi}{|\phi_{23}-\frac{\pi}{2}|} \; &\phi_{23}\le\frac{\pi}{6} \;\; (M\ge m)\\
\frac{\pi}{2|\phi_{23}|} \; &\phi_{23}>\frac{\pi}{6} \;\; (M<m)\\
\end{cases}, &\text{Even Parity}\\
\frac{\pi}{|\phi_{23}-\frac{\pi}{2}|}, &\text{Odd Parity}
\end{cases}
\end{equation*}
while for fermions we have
\begin{equation*}
\kappa_{\rm min}=
\begin{cases}
\frac{\pi}{|\phi_{23}-\frac{\pi}{2}|}, &\text{Even Parity}\\
\begin{cases}
\frac{\pi}{|\phi_{23}-\frac{\pi}{2}|} \; &\phi_{23}\le\frac{\pi}{6} \;\; (M\ge m)\\
\frac{\pi}{2|\phi_{23}|} \; &\phi_{23}>\frac{\pi}{6} \;\; (M<m)\\
\end{cases}, &\text{Odd Parity}
\end{cases}
\end{equation*}
Note that these results do not all immediately reduce to the equal mass
results (i.e. $\phi_{23}=\pi/6$) of Sec.~\ref{SqWell} since the
symmetry at $\phi_{23}$ has not been taken into account here.  That
symmetrization eliminates some solutions, finally giving complete agreement
with the equal mass results.

\section{Recombination for Bosons}
\label{section:recomb_bosons}

\subsection{Cross section and event rate constant}
\label{RateDef}

In the presence of a purely short-range hyperradial potential, the scattering solution for distinguishable particles is of the form
\begin{equation}
\Psi^{dist}\rightarrow e^{ikR\cos{(\phi-\phi')}} + f^{3B}\frac{e^{ikR}}{\sqrt{R}},
\end{equation}
where the ``direction'' of the incident plane wave is parameterized by the angle $\phi'$.  For $\phi'=0$, the incident flux is in
the direction of the first Jacobi vector $x_{12}$, while for $\phi'=\pi/2$, the incident flux is in the direction of $x_{12,3}$.
The quantity $f^{3B}$ represents the elastic three-body scattering amplitude and is given by
\begin{equation}
f^{3B}=\sqrt{\frac{1}{2\pi k}}\sum_{m=-\infty}^{m=\infty}{e^{im(\phi-\phi')}(e^{2i\delta_m}-1)}.
\end{equation}
We have defined the scattering phase shift in the $m$-partial wave as $\delta_m$.
The total integrated cross section is then~\cite{morse_feshbach}
\begin{equation}
\sigma_{dist} = \int_{0}^{2\pi}{d\phi\;|f^{3B}|^2} = \frac{1}{k}\sum_{m=-\infty}^{m=\infty}{|e^{2i\delta_m}-1|^2}
\end{equation}

This expression must be modified to account for three separate issues.  First, we are interested in 
the cross section for an inelastic process.  Second, we must account for the appropriate identical particle symmetry, and 
third, that our asymptotic harmonics are not two-dimensional partial waves.  

Taking these issues into consideration, 
we find --- following the arguments of~\cite{ident_part} --- that the 
recombination cross section for identical bosons in terms of the $S$-matrix is
\begin{equation}
\sigma = \frac{6}{k}\sum_{\kappa,p}{\left|\left<\Phi_{2B,p}^{sym}|\hat{S}|\kappa\right>\right|^2}
\end{equation}
where $\Phi_{2B,p}^{sym}$ denotes the final symmetrized two-body channel function with overall three-body parity 
$p$, and $\kappa$ labels the
three-body entrance channel.  At ultracold temperatures, we need only include the smallest $\kappa=\kappa_{\rm min}$, giving
\begin{equation}
\sigma = \frac{6}{k}\left|\left<\Phi_{2B,e}^{sym}|\hat{S}|\kappa=3\right>\right|^2.
\end{equation}
Similar expressions can be derived for three identical fermions and for systems
with only two identical particles.  In fact, the only change for these other symmetries ---
besides having different dominant $\kappa$ and parity
at threshold --- is the numerical prefactor.

We are not only interested in the cross section, but also in the more experimentally relevant event rate constant,
\begin{equation}
\label{rateK3}
K_3=\frac{\hbar k}{\mu}\sigma= \frac{6\hbar}{\mu}\left|S_{2B,3B}^{sym}\right|^2.
\end{equation}
Here, we explicitly show all factors of $\hbar$ in order to emphasize that this 
quantity has the appropriate units of length$^2$/time.  $K_3$ represents
the probability of a recombination event per atomic triad per unit density of atomic triads per unit time.  
The volume is the full volume spanned by the internal coordinates of the three particle system, so that the 
density is the number of triads per unit two-dimensional volume.  As defined in Eq.~(\ref{rateK3}),
the rate is a function of energy.  To compare with experiment, though, it should be thermally
averaged to give the rate as a function of temperature.  This averaging is especially important 
in 1-D since $K_3(E)$ is not constant at threshold.

\subsection{Zero vs. finite-range interactions}
\label{subsection:zrp_vs_finite}

For identical bosons with $\delta$-function interactions, the elastic 
atom-dimer $S$-matrix element is~\cite{mcguire,thacker,mehta1}:
\begin{equation}
\label{s11_elastic}
S_{11}=\exp{(2i\delta)} = 1 - \frac{24(k_{12,3}a)}{\frac{9}{2}i(k_{12,3}a)^2 + 12(k_{12,3}a)-6i}
\end{equation}
where $k_{12,3}=\sqrt{2\mu_{12,3}(E+B_2)}$.  Simple analysis of Eq.~(\ref{s11_elastic}) shows that $|S_{11}|^2 = 1$ for
all energies, meaning that scattering in the two-body channel is always elastic. Hence, the amplitude for
breakup (and also recombination) is identically zero at all energies.    

\begin{figure}
\begin{center}
\leavevmode
\includegraphics[width=3.375in,clip=true]{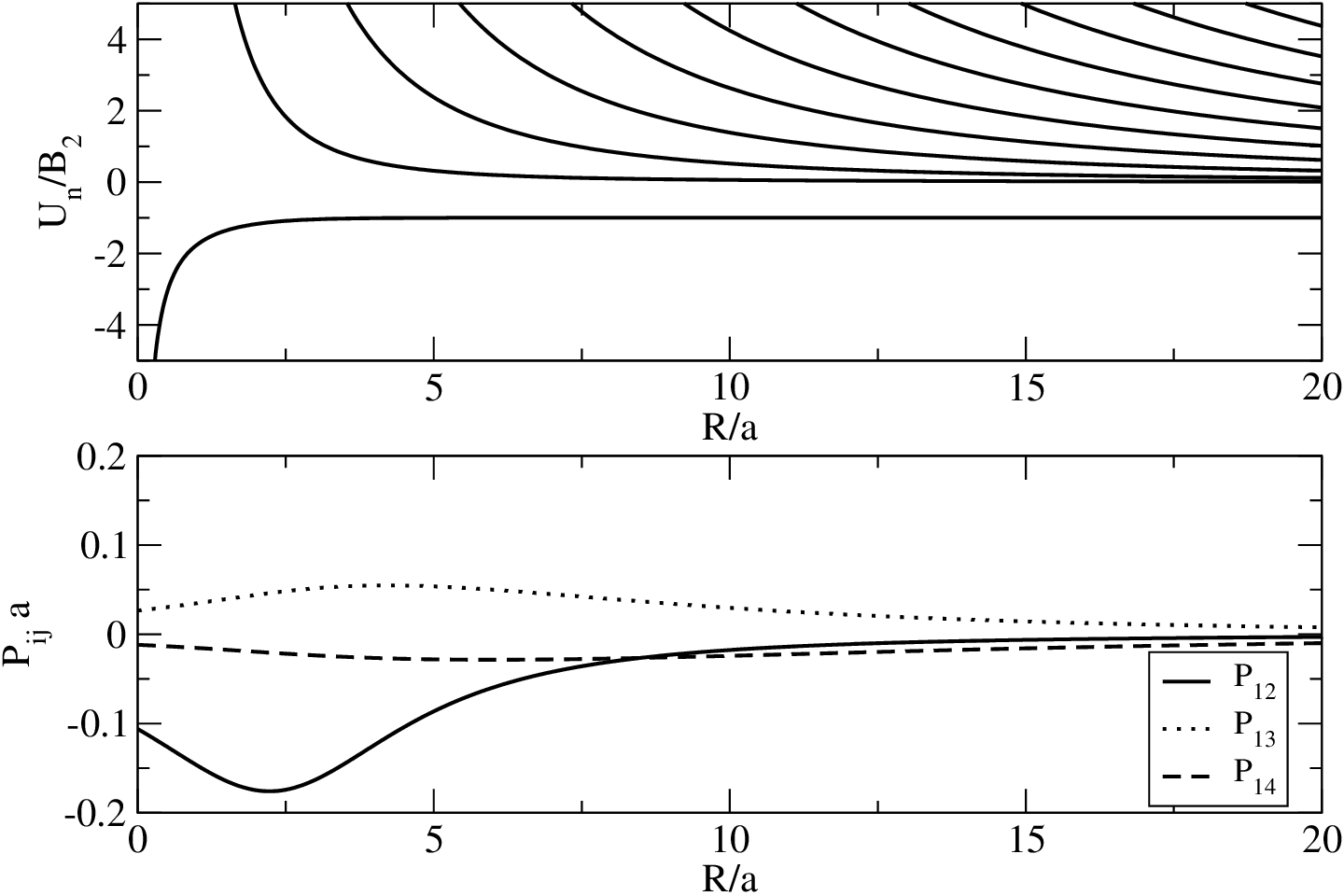}
 \begin{picture}(0,0)(0,0)
      \put(-95,160){\large (a)}
      \put(-95,75){\large (b)}
 \end{picture}
\includegraphics[width=3.375in,clip=true]{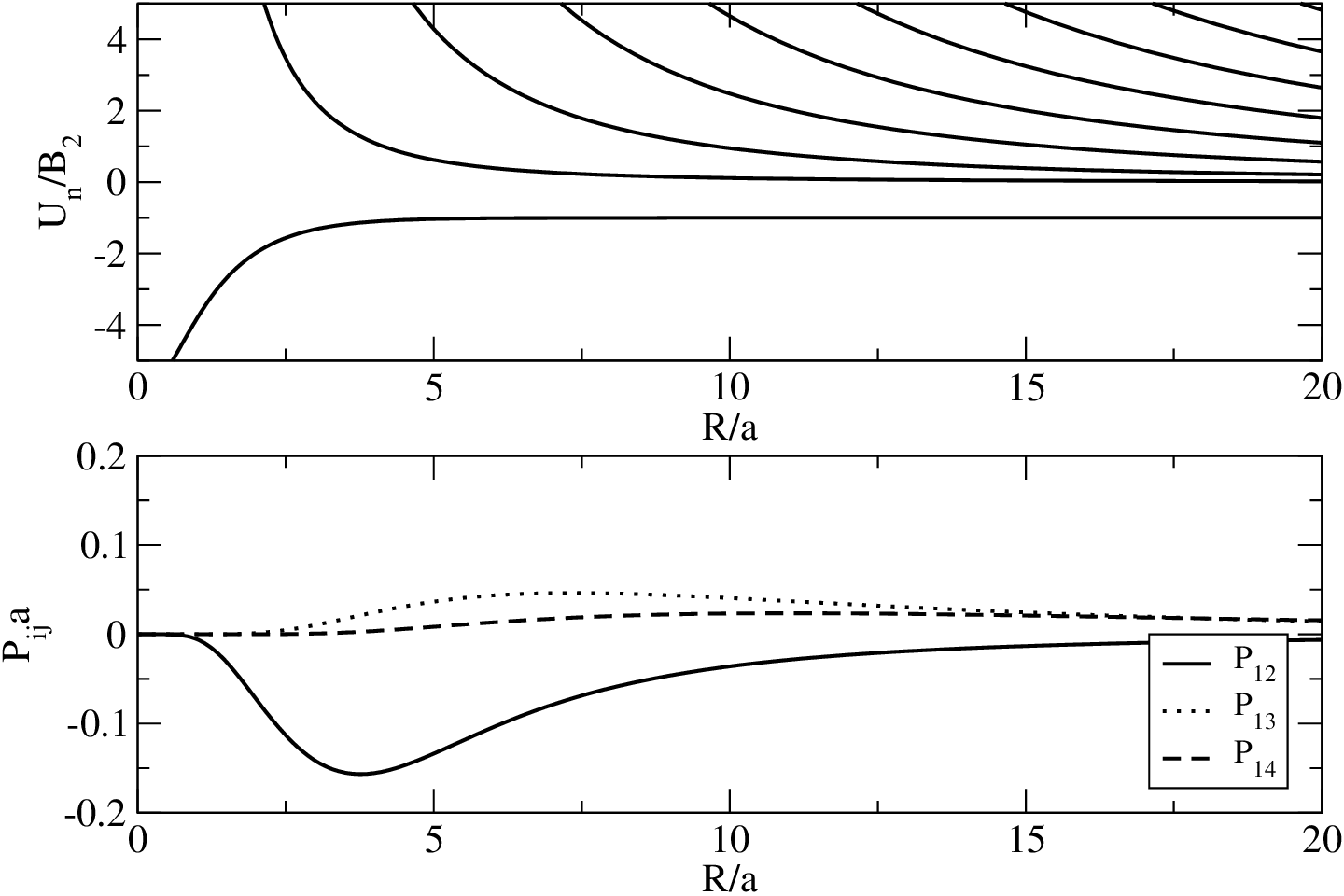}
 \begin{picture}(0,0)(0,0)
      \put(-95,160){\large (c)}
      \put(-95,75){\large (d)}
 \end{picture}
\caption{The adiabatic potential curves (a) and first derivative couplings (b) for $\delta$-function interactions ($a=2$).
The adiabatic potential curves (c) and first derivative couplings (d) for the P\"oschl-Teller potential with $L=a=2$.  All
axes are dimensionless.}
\label{fig:pots_pmat}
\end{center}
\end{figure}
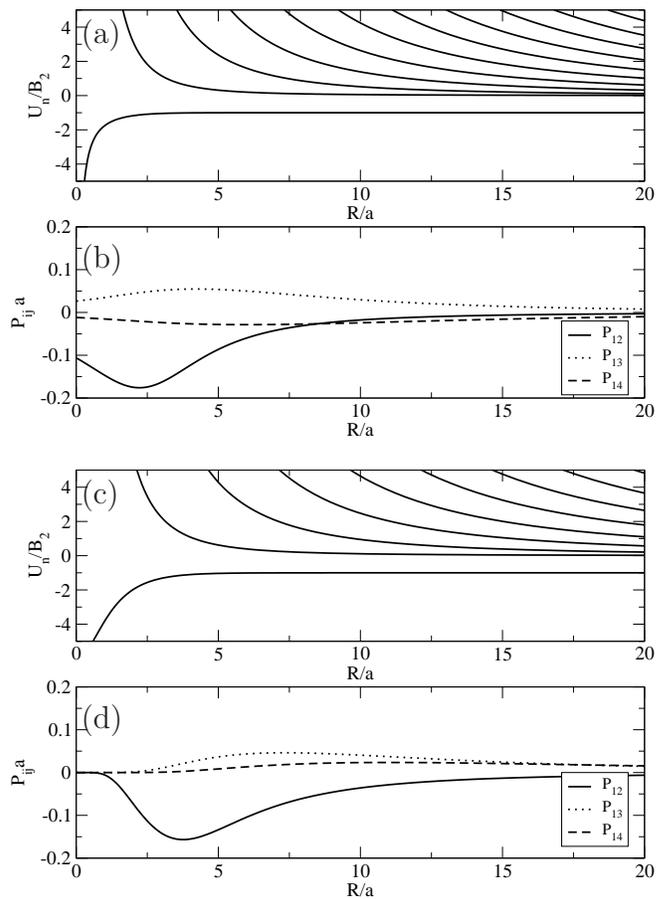

For $\delta$-function interactions, the adiabatic hyperspherical potential curves have previously been 
calculated by solving the transcendental equation Eq.~(\ref{ebdelta})~\cite{gibson_etal,amaya-tapia,mehta1}.  
Since we want to calculate inelastic transition rates, we also require
the nonadiabatic couplings $P_{ij}$ and $Q_{ij}$ in Eqs.~(\ref{pmat}) and~(\ref{qmat}).  In general, the preferred
way to calculate these couplings numerically is from difference
formulas.  It is difficult to obtain high accuracy by differencing, though.
In cases where the adiabatic solution $\Phi$ can be written analytically, however, it is possible to
calculate $\partial\Phi/\partial R$ directly~\cite{kartavstev2006ulp}.   
This is accomplished by differentiating the transcendental equations for $\kappa$ in 
Sec.~\ref{section:threshold} to determine equations for $\partial\kappa/\partial R$.  
Rather lengthy expressions for $P_{ij}$ and $Q_{ij}$ result, but they do allow the couplings to be
calculated essentially exactly.  The first few 
elements of the first row of the antisymmetric matrix $\mathbf{P}$ are shown in Fig.~\ref{fig:pots_pmat}(b) and (d).  The
important feature of this figure is simply that these elements are not zero.  They couple the
two and three-body channels, yet we know from Eq.~(\ref{s11_elastic})
that the amplitude for any process connecting the two and three-body channels must vanish.  
We will numerically demonstrate that despite the nonzero couplings, the solution to the 
coupled radial equations indeed gives vanishing probability for inelastic events.  

In order to facilitate this demonstration, we consider 
the P\"oschl-Teller two-body potential,
\begin{equation}
\label{PT_pot}
V(x)=-D\;{\rm sech}^2{\frac{x}{L}}, ~~~ D>0,
\end{equation}
which gives a Schr\"odinger equation having an analytic solution~\cite{landau1974ctp}.
Defining $\eta=\sqrt{1+4DmL^2}$, 
the zero-energy scattering length for the even parity solution is 
\begin{equation}
a = \frac{L}{2}\left(H_{-\frac{1}{2}-\frac{\eta}{2}} + 
	H_{-\frac{1}{2}+\frac{\eta}{2}}-\pi\sec{\eta\frac{\pi}{2}}\right).
\end{equation}
In this expression, $H_x=\gamma_{EM} + \psi(x+1)$, $\gamma_{EM}$ is the Euler-Mascheroni constant, and
$\psi(x)=\Gamma'(x)/\Gamma(x)$ is the digamma function.  The energy eigenvalues are
\begin{equation}
E=-\frac{\left(2n+1-\eta \right)^2}{4mL^2} \qquad n=0,1,2...
\end{equation}
The scattering length becomes infinite when an even parity 
state sits at threshold, that is, when $E=0$ for $n=0,2,4...$.  

To check whether recombination with the P\"oschl-Teller potential recovers the $\delta$-function result
(\ref{s11_elastic}), we set
$m=1$, fix the scattering length
to $a=2$, and consider the effect of letting $L\rightarrow 0$, approaching the $\delta$-function limit.
When $L=2$ and $D=\frac{1}{2}$, the P\"oschl-Teller potential gives the same two-body binding energy
and scattering length as the 
$\delta$-function.  The only difference is that the former is of finite range.  The difference in the 
resulting three-body potential curves and couplings, however, is more subtle.  
These potentials and couplings are shown in Fig.~\ref{fig:pots_pmat}(c) and (d).
By comparison with the $\delta$-function results in Fig.~\ref{fig:pots_pmat}(a) and (b),
we see that all features for the P\"oschl-Teller results tend to be pushed to
larger $R$.

\begin{figure}
\begin{center}
\leavevmode
\includegraphics[width=3.375in,clip=true]{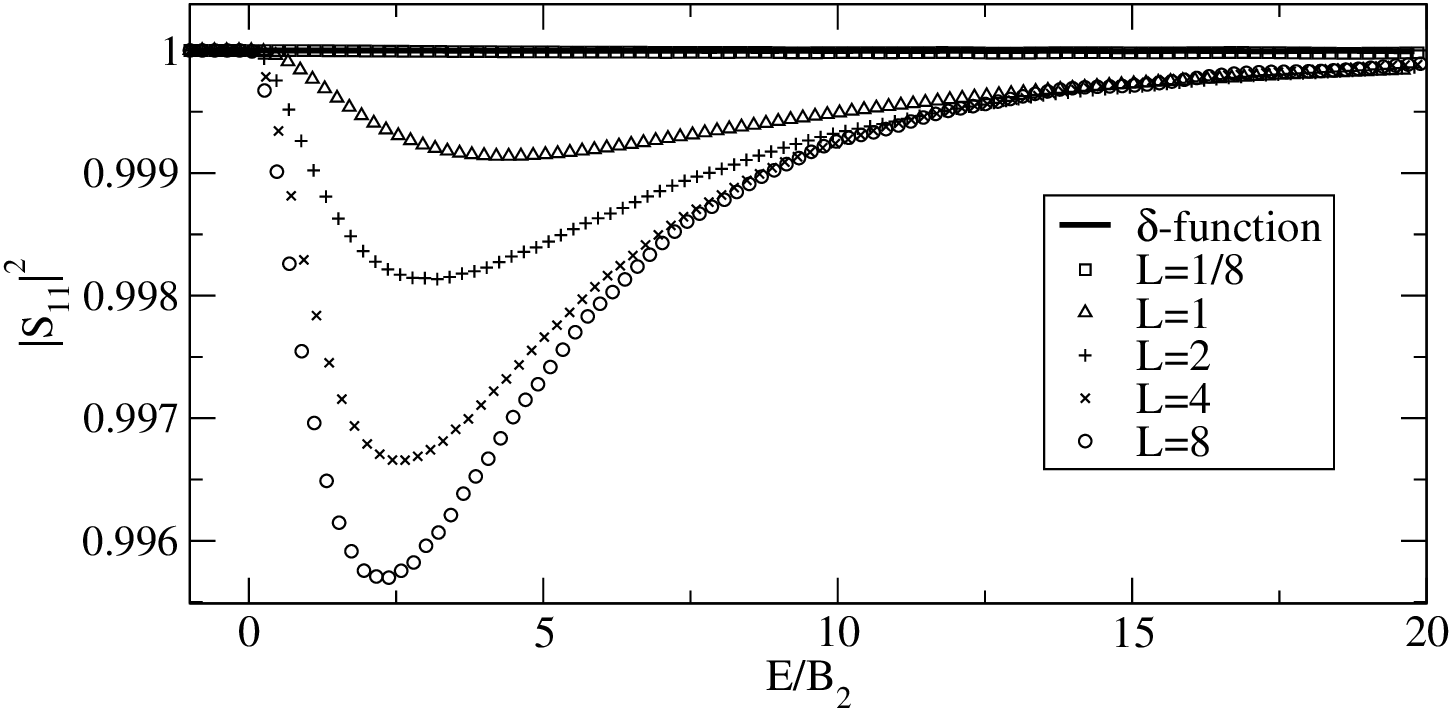}
\caption{The convergence of $|S_{11}|^2$ for finite-range potentials (with range $L$) to the zero-range theory is shown.
The scattering length is held fixed at $a=2$ as $L$ in Eq.~(\ref{PT_pot}) is taken to zero.  
The data for $L=1/8$ (squares) are nearly indistinguishable from
the solid $\delta$-function line.  All axes are dimensionless.
}
\label{fig:ZRP_vs_fin_elastic}
\end{center}
\end{figure}

Our numerically obtained $|S_{11}|^2$ for both kinds of potentials are shown
in Fig.~\ref{fig:ZRP_vs_fin_elastic}.  Below the dimer break-up threshold at $E=0$, the collision is purely elastic.
Above the break-up threshold, the phase-shift in the elastic channel acquires an imaginary part,
which appears as a deviation from $|S_{11}|^2=1$, implying a nonzero probability for break-up of the dimer.  
As indicated in the figure, our numerical calculations 
show that as $L\rightarrow 0$, the collisions do indeed become purely elastic at \emph{all} energies, 
in agreement with Eq.~(\ref{s11_elastic}).
Since the $S$-matrix is symmetric under time-reversal, the amplitude for recombination also vanishes.
It should be stressed that the solid black line in Fig.~\ref{fig:ZRP_vs_fin_elastic} is the result of our numerical calculation
for the $\delta$-function, and not simply a plot the analytical result, Eq.~(\ref{s11_elastic}).

\subsection{Low-energy effective interaction}
\label{subsection:veff_recomb}

While the potential Eq.~(\ref{PT_pot}) has the advantage of yielding an exact analytic two-body solution, it does not facilitate the
study of recombination into a single high-lying two-body state in the limit $a \rightarrow \infty$.  
The disadvantage stems from the fact that the potential is purely attractive.
As is well known, any purely attractive potential in 1-D will support an even parity bound state no matter how small the 
coupling.  Hence, for the potential Eq.~(\ref{PT_pot}) to have $a \rightarrow \infty$ with a single bound state, 
we must let $D \rightarrow 0$.  Therefore, we are motivated to construct a renormalized low-energy potential 
model that supports a single (shallow) bound state with finite couplings in the limit $a \rightarrow \infty$.  We follow the insight of
Ref.~\cite{lepage} and take
\begin{equation}
\label{veff}
V_{eff}(x)=\frac{\Lambda^2}{m}\left[c - 2d + 4d(\Lambda x)^2\right]\exp{(-\Lambda^2 x^2)}.
\end{equation}
The prefactor $\Lambda^2/m$ gives the interaction units of energy in atomic units, and $m$ is the mass of the helium atom in
atomic units ($m=7296.299 m_e$).  One can tune the free couplings $c$, and $d$ so that the 
potential reproduces physically reasonable values for the
scattering length and effective range of, for example, $^4$He~\cite{janzen_aziz}: 
$a=208 \; \text{a.u.}$ and $r_0=14 \; \text{a.u.}$.  A reasonable requirement of any 
renormalized model is that low-energy observables remain independent of the momentum cutoff $\Lambda$.
We find that the values $c=1.97151$ and $d=-1.46165$ for $\Lambda=0.16$~a.u. yield the desired
effective range parameters.  These values also give a two-body binding $B_2=3.39985\times 10^{-9} \; \text{a.u.}$.  We have verified
that $B_2$ varies by less than one percent over a wide range of cutoffs $0.08 \le \Lambda \le 0.3$, and that the 
couplings $c$ and $d$ are of order unity over this range.  

\subsection{Low-energy scaling behavior}

\begin{figure}
\begin{center}
\leavevmode
\includegraphics[width=3in,clip=true]{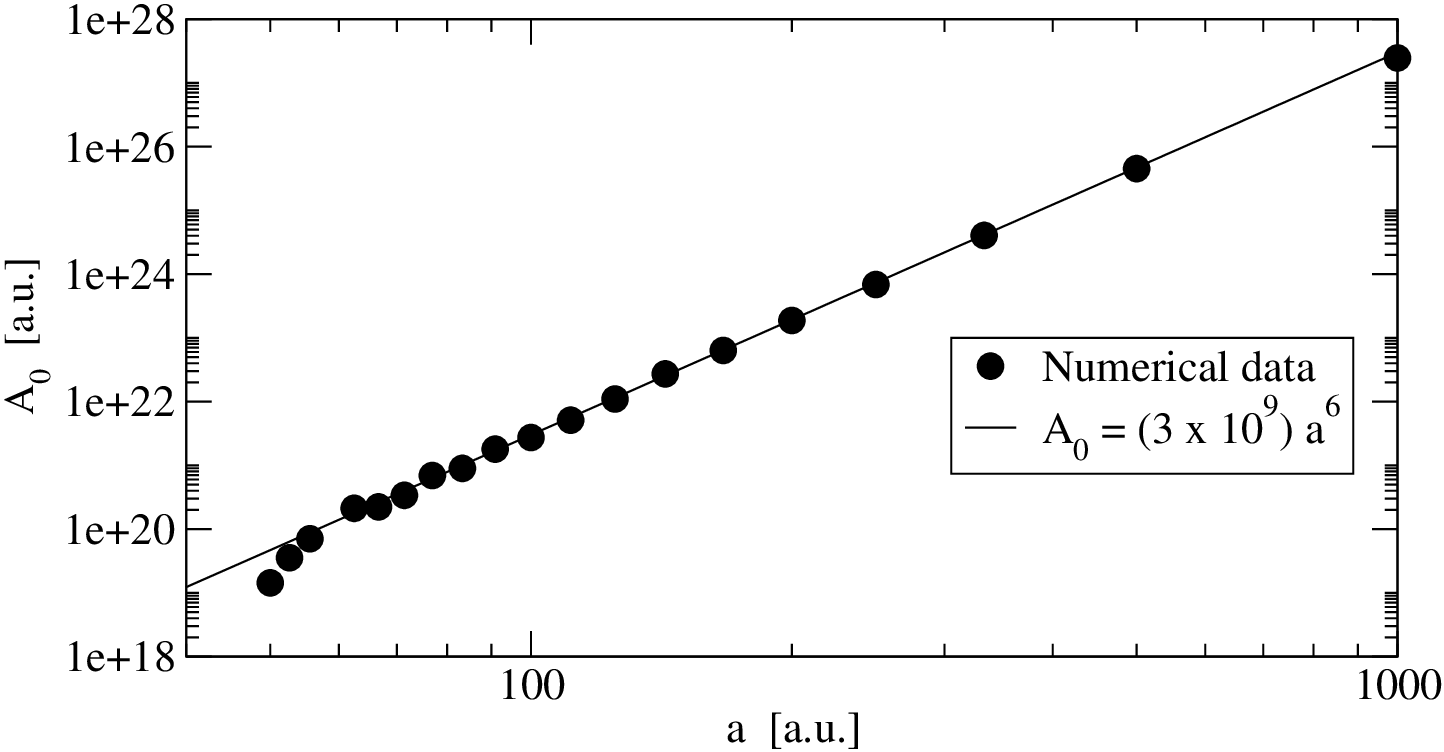}
\caption{A numerical demonstration of $a^6$ scaling of the three-body recombination probability is shown.  The
black line is $A_0 = 3.0 \times 10^9 a^6$, and the circles are our numerical results.  The units of $A_0$ are atomic 
units of inverse energy cubed consistent with the use of Eq.~\ref{veff} to model the $^4$He two-body interaction.
}
\label{fig:asixth_scaling}
\end{center}
\end{figure}

From Eq.~(\ref{S_scaling}) and Sec.~\ref{section:threshold}, we know that
the inelastic transition probability must scale as $|S_{12}|^2\sim k^6 \sim E^3$ at threshold
since $\kappa_{\rm min}=3$ for three bosons.  This behavior is already plausibly seen in
Fig.~\ref{fig:ZRP_vs_fin_elastic} since this scaling for $S_{12}$ implies $1-|S_{11}|^2\sim E^3$ at
the three-body breakup threshold.
When the scattering length $a$ is the longest length scale, we also know from Eq.~(\ref{S_a_scaling}) that
$|S_{12}|^2\sim(ka)^6 \sim E^3a^6$.  
We demonstrate this behavior quantitatively in Fig.~\ref{fig:asixth_scaling}.
The points in the figure were obtained by first tuning the effective potential Eq.~(\ref{veff}) to reproduce
a given scattering length, and then calculating the recombination probability near threshold, $E\rightarrow 0$. 
Finally, we divided the probability by $E^3$ to extract
the constant of proportionality.  The constant 
$A_0$ in $|S_{12}|^2=A_0 E^3$ is plotted as a function of the scattering length in Fig.~\ref{fig:asixth_scaling}.  
The $a^6$ scaling is 
clearly seen when compared to the solid line which is $A_0= 3.0\times 10^9  a^6$.

It is interesting to note that the lack of inelastic processes for zero-range interactions is actually a consequence of
perfect destructive interference in the exit channel.  Indeed, since the couplings $P_{ij}$ and $Q_{ij}$ are non-zero, the only
way for the inelastic probability to vanish is through some sort of interference effect.  It is possible to demonstrate this
perfect interference by adding an arbitrary short-range three-body interaction $V_{3}(R)$ of characteristic length $L_3$ 
to the zero-range hyperradial potentials.  The short-range three-body interaction destroys the perfect interference and leads to a 
nonzero recombination probability.  Considering the ratio $C=\frac{1-|S_{11}|^2}{(ka)^6}$ near threshold, 
we find a nonuniversal power law (that 
depends on the short-range nature of $V_3$) of the form $C \sim (L_3/a)^{c_1}$.  Although the particular value of $c_1$ is
nonuniversal, we always find $c_1 > 0$ such that $K_3 \rightarrow 0$ as $L_3 \rightarrow 0$.

\section{Universality in One Dimension}
\label{1Duniversality}

In three dimensions, the three-body problem exhibits universal features in the limit
$|a|\rightarrow \infty$ related to Efimov physics~\cite{efimov1,dincao2005mee,nielsen_etal,braaten2006ufb}.  Here, 
we consider this limit in one dimension. 
For $L\ll R \ll |a|$ --- with $L$ generally the characteristic size of the two-body 
interaction --- we find that
\begin{equation}
q^2 =-\kappa^2 \approx \frac{12 \alpha R}{\sqrt{3}\pi a}
\end{equation}
to leading order in $R/a$
for $a>0$ and $a<0$, respectively.  The corresponding adiabatic potentials are
\begin{equation}
U(R) = \frac{\gamma_0}{2\mu a R}
\label{UnivPot}
\end{equation}
with
\begin{equation}
\gamma_0 = -\frac{12}{3^{1/4}2^{1/2}\pi} \approx -2.052277.
\label{gamma0}
\end{equation}
We thus find an attractive, universal $R^{-1}$ potential when $a>0$, and a
similarly universal, repulsive $R^{-1}$ potential when $a<0$.  In the former
case, this universal potential converges to the highest-lying two-body threshold
while for the latter, it is the lowest potential converging to the three-body
breakup threshold.

This result can be derived in several ways.  For instance, the $\delta$-function
transcendental equations Eqs.~(\ref{delta_tran_2b}) and 
(\ref{ebdelta}) for positive and negative scattering lengths, respectively,
can be solved in the small $q$ or $\kappa$ limit.  The exact same universal curves
can similarly be extracted from the analogous quantization equations for 
the square-well interaction [Eq.~(\ref{EvenBosonSqWell}), for example]
using 
\begin{equation}
\beta=\left(\frac{4}{3}\right)^\frac{1}{4} \frac{R}{L} n\pi\left( 1+\frac{L}{n^2\pi^2 a}\right)
, \qquad n>0,
\label{Large_a_beta}
\end{equation}
assuming there are $n$ two-boson bound states.  This choice for $\beta$ follows
from the fact that the two-body square-well has a zero-energy bound state when
$\sqrt{2\mu_{2B}V_0}L=n\pi$.  Reducing this phase slightly gives $a<0$; and increasing
it, $a>0$.  Finally, the universal curves can be obtained quite generally from 
the quantization conditions for arbitrary short-range potentials, Eq.~(\ref{GenInt}),
using this expression for the log-derivative in the limit $2\mu_{2B}EL^2 \ll 1$:
\begin{equation}
b = \frac{1}{a-L}.
\end{equation}

It is clear that the universality of the 1-D problem has a different character than the
3-D problem.  First, $\gamma_0$ is the coefficient of $R^{-1}$ instead of $R^{-2}$ as in 3-D.  Second, 
the universal potential Eq.~(\ref{UnivPot}) actually vanishes in the limit $|a|\rightarrow\infty$.
The lowest three-body continuum potential in this case is instead an attractive $R^{-3}$ potential.
For square-well interactions, we can derive it explicitly using $\beta$ from Eq.~(\ref{Large_a_beta})
with $|a|\rightarrow\infty$ in Eq.~(\ref{EvenBosonSqWell}), yielding
\begin{equation}
U(R) = -\left(\frac{3}{4}\right)^\frac{1}{4} \frac{n^2 \pi L}{2\mu R^3}.
\label{Inf_a_U}
\end{equation}
This potential, however, is not universal, although numerical calculations with four different
short-range two-body potentials suggest that two aspects are general: (i) the attractive $R^{-3}$
behavior and (ii) the increasing interaction strength with $n$.

As $|a|\rightarrow \infty$, the solution in the lowest channel in the 
region $L \ll R\ll |a|$ will obey the equation
\begin{equation}
\label{univ_se}
\left[-\frac{1}{2\mu R}\diff{}{R}\!\left(\!R\diff{}{R}\right) \!+\! \frac{\gamma_0}{2\mu a R} \!-\! \frac{Q_{11}(R)}{2\mu}\right]\!F(R)=E F(R).
\end{equation}
The diagonal nonadiabatic coupling $-Q_{11}(R)$ is always repulsive and falls off as $R^{-2}$, while the universal $R^{-1}$ 
term will vary smoothly from an attractive to a repulsive potential as $a$ varies from $+\infty$ to $-\infty$.  
When $|a|=\infty$, of course, the $R^{-1}$ potential vanishes, leaving just the $R^{-3}$
potential in Eq.~(\ref{Inf_a_U}).

\begin{figure}[!t]
\begin{center}
\leavevmode
\includegraphics[width=3.375in,clip=true]{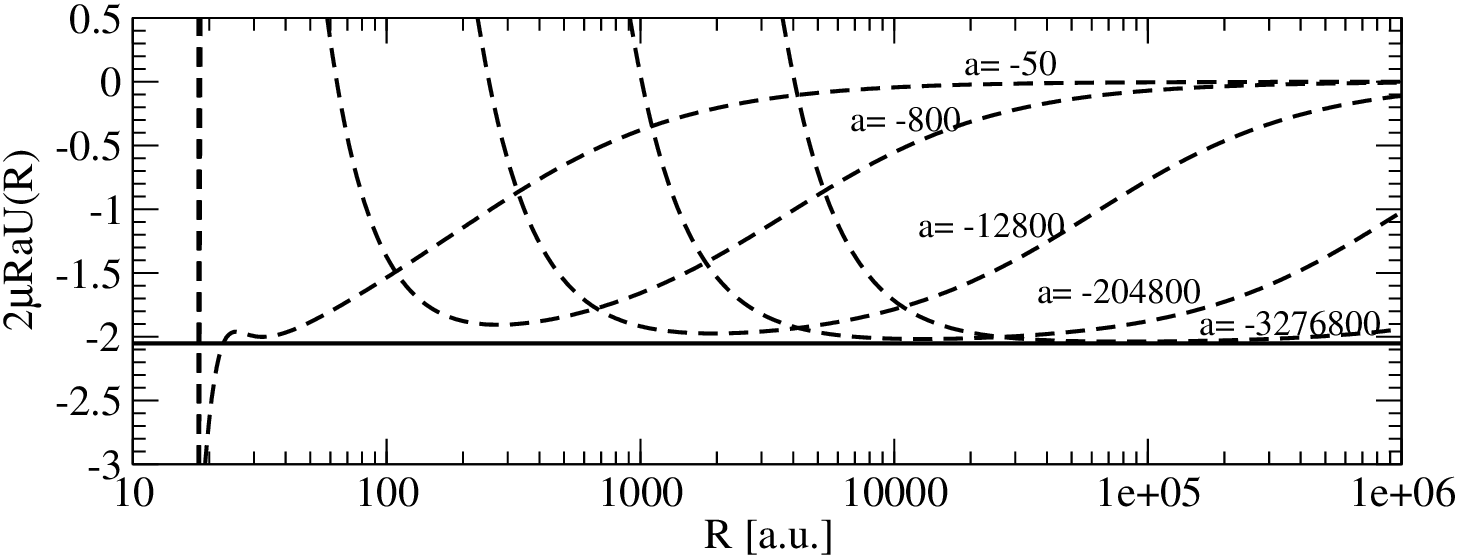}
\caption{The lowest adiabatic hyperspherical potential curve for the two-body interaction in Eq.~(\ref{veff})
with several different values of $a<0$ ($a$ given in a.u.).  The curve is multiplied by the factor $2\mu R a$ to give
a unitless quantity and more clearly reveal where the potential behaves as a repulsive $R^{-1}$.  The horizontal solid line is the
universal number $\gamma_0 \approx -2.05$~(\ref{gamma0}).
}
\label{fig:a_lt_0_univ}
\end{center}
\end{figure}

In order to demonstrate that $\gamma_0$  is indeed a universal constant, we again turn to the effective potential Eq.~(\ref{veff}).
We have calculated three-body potential curves Eq.~(\ref{veff})
giving different two-body scattering lengths (the effective range is held constant at $14$~a.u.).  
Figure~\ref{fig:a_lt_0_univ} shows the lowest three-body potential for increasingly negative scattering lengths.
The potential curves have been multiplied by the
factor $2\mu a R$ in order to more clearly reveal the universal behavior.  
We see that the curves do in fact approach $\gamma_0$ over a range of $R$
consistent with the condition $r_0\ll R\ll |a|$.  At $R\gg |a|$, the
potentials again approach the three-body breakup threshold with the $R^{-2}$
behavior predicted in Sec.~\ref{section:threshold}
(which translates to $R^{-1}$ as plotted in the figure).

\begin{figure}[!t]
\begin{center}
\leavevmode
\includegraphics[width=3.375in,clip=true]{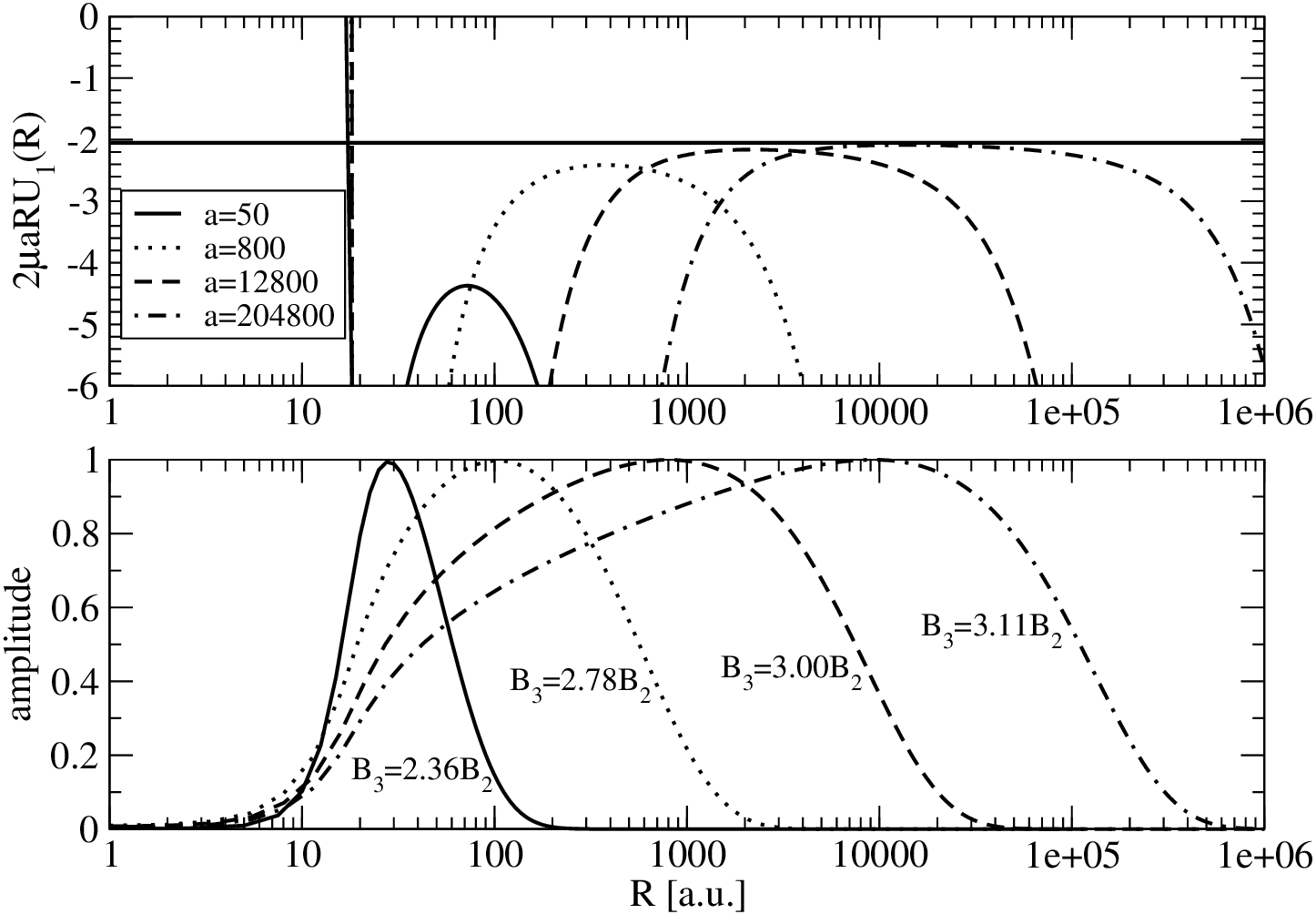}
 \begin{picture}(0,0)(0,0)
      \put(-97,165){\large (a)}
      \put(-97,80){\large (b)}
 \end{picture}
\caption{(a) The lowest adiabatic hyperspherical potential curve for the two-body interaction in Eq.~(\ref{veff})
with several different values of $a>0$ (units of $a$ are a.u.).  The curve is multiplied by the factor $2\mu R a$ to
reveal the region in which the potential behaves as an attractive $R^{-1}$.  The horizontal solid line is the
universal number $\gamma_0 \approx -2.05$~(\ref{gamma0}).  (b) The hyperradial wave functions for each
potential in (a).  The wave functions have been scaled such that their maximum is unity.  As $a \rightarrow \infty$, 
we expect the three-body binding $B_3 \rightarrow 4 B_2$, as is the case for $\delta$-function interactions~\cite{mcguire}. 
}
\label{fig:univ_boundstates}
\end{center}
\end{figure}
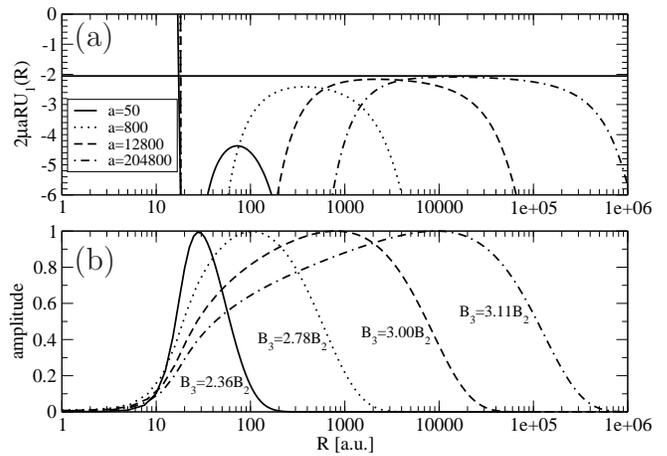

For large positive scattering lengths, the lowest potential curve --- the two-body channel --- 
supports a three-body bound state
so long as there is at most one weakly bound two-body state.  
Figure~\ref{fig:univ_boundstates} shows the lowest potential curve for various values of the two-body scattering length along
with the hyperradial wave functions for the three-body bound state in this channel. 
It is clear that the universal $R^{-1}$ portion
of the curve has a strong influence on these states since
a significant portion of the necessary phase is accumulated in this region.  
This argument is supported by a simple WKB calculation using the 
the Bohr-Sommerfeld quantization condition 
\begin{equation}
\int_{R_1}^{R_2}{dR\; \sqrt{2\mu E -\frac{\gamma_0}{aR}}} = \left(n+\frac{1}{2}\right)\pi
\label{WKB_bound}
\end{equation}
for classical turning points $R_i$.  We note that the Langer correction does not appear in this
equation since it cancels the attractive $R^{-2}$ term obtained from eliminating the first derivative
in the hyperradial kinetic energy --- which is required to use this WKB phase integral.
The energies of the nodeless solutions as $a\rightarrow \infty$ using the above equation are 
in reasonable agreement with a B-spline calculation using the numerical
potential curves.  These results are tabulated in Table~\ref{table:wkb}.

Further examination of Eq.~(\ref{WKB_bound}) also shows why --- despite the
long-range $R^{-1}$ behavior --- there is only a single three-body bound state.
If we evaluate the WKB phase at the two-body threshold for $R$ between $L$ and
$a$, then we find that it lies between $\pi/2$ and $0.817\pi$
for all $a/L$ between 8.07286 and infinity.
The phase remains finite because the strength of the $R^{-1}$ potential
decreases with $a$ at the same time that the domain over which it holds grows with $a$.
So, there is sufficient phase accumulated in this universal region alone to support
a single bound state by Eq.~(\ref{WKB_bound}).  Moreover, the phase contributed
from the small-$R$ region, $R\le L$, would have to exceed roughly $0.7\pi$ to 
produce a second bound state.  While this is not impossible, it does not seem likely.

\begin{table}[!t]
\caption{\label{table:wkb}
The three-body bound-state energies 
are shown in atomic units for various two-body potentials of the form Eq.~(\ref{veff}) tuned to
give the scattering lengths in the first column.  The WKB estimates improve as $a\rightarrow \infty$.}
\begin{center}
\begin{tabular}{ccc}
\hline\hline
$a$ & Numerical B-spline & WKB \\
\hline
$50$ & $-1.8644\times 10^{-7}$ & $-4.8391\times 10^{-8}$\\
$800$ & $-6.0569\times 10^{-10}$ & $-4.8839\times 10^{-10}$\\
$12800$ & $-2.5156\times 10^{-12}$ & $-2.6829\times 10^{-12}$ \\
$204800$ & $-1.0324\times 10^{-14}$ & $-1.1533\times 10^{-14}$ \\
\hline\hline
\end{tabular}
\end{center}
\end{table}

\begin{figure}[!t]
\begin{center}
\leavevmode
\includegraphics[width=3.375in,clip=true]{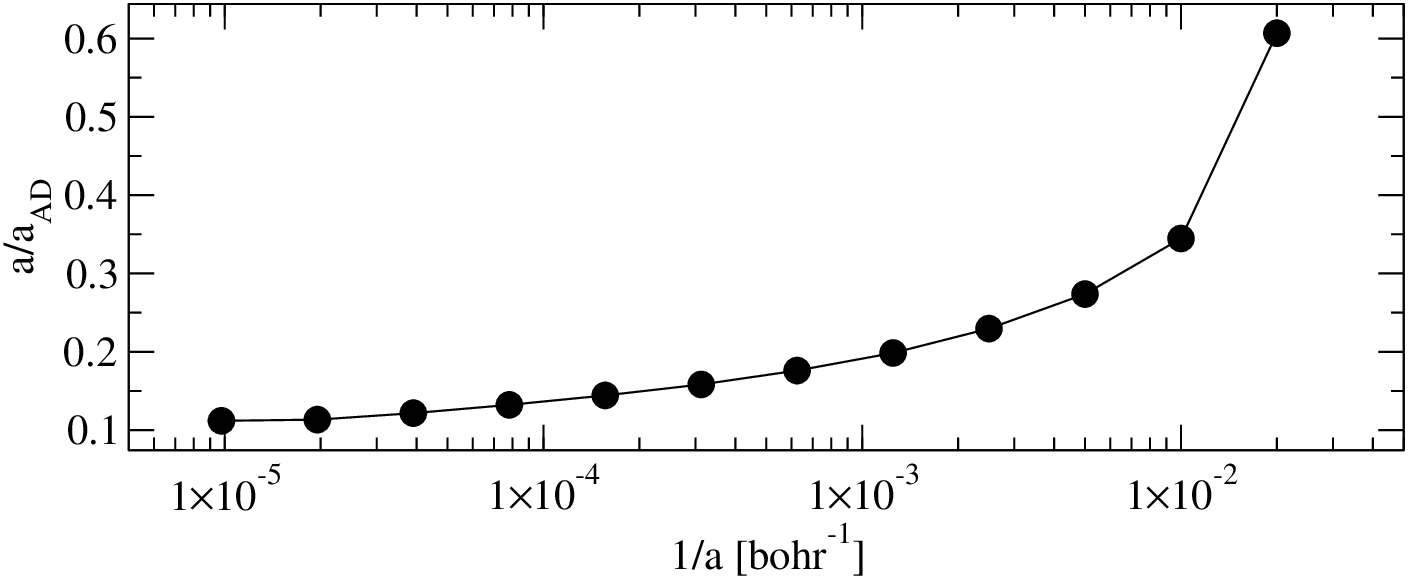}
\caption{The inverse of the atom-dimer scattering length multiplied by the atom-atom scattering length
is shown as a function of the inverse of the atom-atom scattering length.  In the limit that $a\rightarrow \infty$, we find
}
\label{fig:atom_dimer_scatlen}
\end{center}
\end{figure}

Finally, we consider the atom-dimer scattering length $a_{AD}$ in the universal limit.  Again using Eq.~\ref{veff}, we calculate
$a_{AD}$ as $a \rightarrow \infty$.  Our results (plotted in Fig.~\ref{fig:atom_dimer_scatlen}) show a clear
trend towards $a_{AD} \approx 10 a$.  This finding is consistent with the presence of a bound-state at the atom-dimer 
threshold in the $\delta$-function model~\cite{mcguire,amaya-tapia,mehta1}.  As $a \rightarrow \infty$ this zero-energy resonance
results in $a_{AD} \rightarrow \infty$.

Before we end our discussion of universality, it is worth mentioning the consequences
for the recombination rate.  So long as $a$ is finite, the lowest three-body continuum 
potential still behaves as predicted in Sec.~\ref{section:threshold} for $R\gg|a|$.
The power-law scaling of the $S$-matrix with $E$ and $a$ is thus the same as found in that section.
The $a$-dependence is modified, though, by the universal region of the 
potential.  Using arguments similar to those in Refs.~\cite{esry1999rta,dincao2005sls}, we
can use WKB to determine the modifications.  For $a<0$, the coupling driving recombination
peaks around $R=L$, so the system must tunnel through the repulsive $R^{-2}$ potential
at $R\gg|a|$ and through the repulsive $R^{-1}$ universal in the region $L\ll R\ll|a|$
in order to recombine.  The WKB tunneling integral then leads to the modified threshold
scaling for $a<0$
\begin{equation}
\label{suppression1}
K_3 \propto \frac{\hbar}{\mu}(ka)^6 \exp\left[{-4 \sqrt{|\gamma_0|}\left(1-\xi \sqrt{\frac{\! L}{|a|}}\right)}\right],
\end{equation}
where $\xi$ is a numerical constant on the order of unity that depends on the exact range of $R$
over which the universal potential is valid.

For $a>0$, in analogy to the 3-D case, recombination can be modified by interference
to give
\begin{equation}
\label{interference}
K_3 \propto \frac{\hbar}{\mu}(ka)^6 \sin^2 \left[2 \sqrt{|\gamma_0|}\left(1-\xi \sqrt{\frac{\! L}{|a|}}\right)+\Phi\right]
\end{equation}
where $\Phi$ is the short-range, $R\ll L$, phase accumulated in the non-universal portion
of the two-body channel.

When $|a|=\infty$ and the incident three-body continuum potential is given by Eq.~(\ref{Inf_a_U}),
the threshold scaling is nontrivially modified.  The attractive $R^{-3}$ potential is
equivalent to $\kappa_{\rm min}=0$, so the recombination rate is, in fact, independent
of energy at threshold, $K_3\propto {\rm const}$.  It turns out that odd parity, identical
fermions share this threshold law in the limit that the two-body \emph{odd}-parity partial wave scattering
length --- appropriate for fermion-fermion interactions --- goes to infinity.  
The threshold scaling for odd parity bosons and even parity fermions
is also changed from the predictions of Sec.~\ref{section:threshold}.  In the limit that their
respective scattering lengths are infinite, $K_3\propto E^3$ for both cases.

\section{Discussion and Summary}
\label{discussion}

The work presented here has been carried out in strictly one dimension.  It is worth commenting on
the relation of this study to experimentally-realizable, effective 1-D geometries.  Olshanii~\cite{olshanii1998asp} has determined the 
effective 1-D scattering amplitude for two particles with 3-D $s$-wave scattering length $a_{3D}$ interacting under 
strong cylindrical harmonic confinement.  His analysis leads to the following one-dimensional effective range expansion:
\begin{equation}
\label{olshanii_ert}
k\tan{\delta} = \frac{1}{a} + \frac{\zeta(3/2) a_\perp^3}{8 a^2}\frac{1}{2}k^2 + O(k^4),
\end{equation}
where $a_\perp$ is the oscillator length in the confined direction, and $\zeta(x)$ is the Riemann zeta function.  
The 1-D scattering length is thus determined at zeroth order in $k$ to be 
\begin{equation}
a= -\frac{a_\perp^2}{2a_{3D}}\left(1+\zeta(1/2)\frac{a_{3D}}{a_\perp}\right).
\end{equation}

One may argue that the 3-D effective range should also be present in the $k^2$ term of
Eq.~(\ref{olshanii_ert}), 
but since $a_{\perp} \gg r_{0}^{3D}$ in general, it is reasonable to assume that the 
contribution from $r_0^{3D}$ is
small compared to $a_\perp$.  With this approximation in hand, it is possible to calculate the 
3-D parameters $a_\perp$ and $a_{3D}$ that correspond to the 1-D parameters $a$ and $r_0$ that
we have used.  For $r_0=14$ a.u. and our larger values
of $a$, we find experimentally accessible values for $a_\perp$ and $a_{3D}$.
All values are tabulated in Table~\ref{table:3dparams}.  One notable point is that the 3-D scattering length is negative for 
positive 1-D scattering lengths.  So while there is no shallow two-body bound-state in the 3-D system, a shallow bound-state
appears as a result of the cylindrical confinement.

\begin{table}
\caption{\label{table:3dparams}
The 3-D parameters $a_{3D}$ and $a_{\perp}$ are calculated to give the desired 1-D effective range $r_0=14$ a.u. and 
the 1-D scattering length in the first column.}
\begin{center}
\begin{tabular}{crr}
\hline\hline
$a$ (a.u.) & ~~$a_{3D}$ (a.u.) & ~~$a_\perp$ (a.u.) \\
\hline
$50$ & $-73.66$ & $47.50$ \\
$200$ & $-63.62$ & $119.70$ \\
$800$ & $-78.46$ & $301.62$ \\
$12800$ & $-160.84$ & $1915.14$ \\
$204800$ & $-377.39$ & $12160.40$ \\
$3276800$ & $-925.65$ & $77213.80$ \\
\hline\hline
\end{tabular}
\end{center}
\end{table}

Finally, we comment on the role of three-body recombination in one recent experiment.  
The experiment by Tolra {\it et al}~\cite{tolra_etal} has
measured three-body recombination rates in order to probe properties of the many-body wavefunction.  This idea was originally 
proposed by Kagan {\it et al}~\cite{kagan1985ebc}, who showed that the event rate constant $K_3$ is proportional to 
the three-body local correlation function $g_3$.  Therefore a measured reduction in $K_3$ from a 3-D system
to a 1-D system in Ref.~\cite{tolra_etal} was interpreted as a reduction in $g_3$ and a clear signature of enhanced correlations.

While a complete description of three-body recombination in atomic waveguides requires a full-scale calculation of three particles in 
a confinement potential, we attempt to address some of the issues within our 1-D framework.  
Our work suggests an alternative explanation of the observed suppression of $K_3$ in terms of the three-body hyperradial potentials.  
For the experimental parameters of Ref.~\cite{tolra_etal} ($a_{3D}\approx 100$~a.u., $a_{\perp}\approx 1100$~a.u.), 
Olshanii's formula Eq.~(\ref{olshanii_ert}) gives a large negative 1-D scattering length $a \approx -5300$~a.u. indicating the absence of
a shallow 1-D bound state.  The wavefunction in the lowest three-body entrance channel must therefore tunnel under both the
repulsive $\kappa_{min}^2/(2\mu R^2)$ potential in the region $R \gg |a|$ and the repulsive $\gamma_0/(2\mu a R)$ (recall that $\gamma_0$ 
and $a$ are both negative) potential in the region
$a_{\perp} \ll R \ll |a|$ before reaching the region $R < a_{\perp}$ where a recombination event into a deep two-body channel may 
occur.  Therefore the measured suppression in the confined geometry could be the combined effect of 
the threshold law $K_3^{1D} \propto \frac{\hbar}{\mu} (ka)^6$ and the universal $R^{-1}$ 
barrier given in Eq.~(\ref{UnivPot}) leading to the suppression given in Eq.~(\ref{suppression1}).

While we are confident that we have solved the 1-D Schr\"odinger equation accurately, it is difficult to make quantitative
claims regarding the measured suppression~\cite{tolra_etal}.  One difficulty arises from the fact that the temperature of the system in 
Ref.~\cite{tolra_etal} is not very well characterized, and therefore, it is unclear what energy three-body collisions occur at.  
In addition, there are a number of theoretical complications which can only be accounted for by performing a full-scale 
calculation of three interacting bosons in a confined geometry.  For instance, since there are excited radial modes of the trap and the 
two-body binding energy is typically much larger than the spacing of these radial modes, there will be a series of two-body thresholds
attached to each excited radial mode.  All of these thresholds are open at the initial three-body threshold energy, and
our calculations have not accounted for this.  Also, our strictly
1-D calculations undoubtedly do not correctly represent the hyperradial potentials 
in the region $R \lesssim a_\perp$ where the system is neither 
purely 3-D nor strictly 1-D in nature.  Since the potential appears in the exponent of the WKB tunneling integral, any uncertainty in the
potential leads quantitatively to very different results for the suppression of $K_3$.  Nevertheless, in view of our findings regarding
universality in 1-D systems, it is possible that the observed suppression of $K_3$ in Ref.~\cite{tolra_etal} was a direct probe of the
universal $R^{-1}$ potential Eq.~(\ref{UnivPot}).

We have shown in this study how inelastic processes such as three-body recombination and collision induced break-up behave
near threshold for all combinations of identical particles.  For the system of three
identical bosons, we have further investigated the behavior of these processes
with respect to the range of the pair-wise interactions and to the two-body scattering length.  We have demonstrated 
numerically that the probability for inelastic events vanishes as the range of the interaction is taken to zero 
in agreement with previous analytic results~\cite{mcguire}.  
Our work is related to a recent Letter that has shown 
how inelastic processes could in fact be possible if one considers a zero-range two-channel model
with a confinement induced Feshbach resonance~\cite{yurovsky_etal}.  Finally we have explored in detail the nature of universality for
three-body systems in 1-D.

\begin{acknowledgments}
We would like to thank D.~Blume and B.~Granger for useful discussion during the early stages of this work.  
NPM would also like to thank M.~Olshanii, Z.~Walters and S.~Rittenhouse for useful discussions.  
We acknowledge support from the National Science Foundation.
\end{acknowledgments}

\appendix
\section{Symmetry Operations}
\label{app:symmops}

The pair permutation operators $P_{ij}$ have the following effects on the hyperangle:
\begin{align}
P_{12}\phi &= \pi-\phi \nonumber\\
P_{23}\phi &= \frac{\pi}{3}-\phi \nonumber\\
P_{31}\phi &= \frac{5\pi}{3}-\phi. 
\end{align}
Solutions of definite parity may be found by via the operation
\begin{equation}
\Pi\phi = \pi + \phi.
\end{equation}
For three identical bosons, the symmetric projection requires
\begin{equation}
{\cal S}=1+P_{12}+P_{23}+P_{31}+P_{12}P_{23}+P_{12}P_{31}.
\end{equation}
Similarly, for three identical fermions, the antisymmetric projection operator is 
\begin{equation}
{\cal A}=1-P_{12}-P_{23}-P_{31}+P_{12}P_{23}+P_{12}P_{31}.
\end{equation}

To determine the boundary conditions required by permutation and 
parity, it is useful to apply these operators to the free particle solutions
of Eq.~(\ref{AdEqn})~\cite{suno2002tbr}.  Doing so, we find that the simultaneous symmetry 
eigenstates are real harmonics with indices that are multiples of three.
Explicitly, for bosons
\begin{equation}
\Phi(\phi) \propto 
\begin{cases}
\cos{3n\phi}, ~~~ n=0,2,4,\ldots, &\text{Even Parity}\\
\sin{3n\phi}, ~~~ n=1,3,5,\ldots, &\text{Odd Parity} 
\end{cases}
;
\label{FreeIdentBosons}
\end{equation}
and for fermions,
\begin{equation}
\Phi(\phi) \propto
\begin{cases}
\sin{3n\phi}, ~~~ n=2,4,6,\ldots, &\text{Even Parity} \\
\cos{3n\phi}, ~~~ n=1,3,5,\ldots,  &\text{Odd Parity} 
\end{cases}
.
\label{FreeIdentFermions}
\end{equation}
From these expression, we conclude first of all that we can reduce
the integration range by a factor of 12 --- as expected --- from $2\pi$
to only $\pi/6$ in $\phi$.  We can also
conclude, for instance, that three identical fermions in
the odd parity state must obey the boundary conditions $\Phi'(0)=0$ and $\Phi(\pi/6)=0$.

If there are only two identical particles, then
we need the operators ${\cal S}=1+P_{12}$ for identical bosons and
${\cal A}=1-P_{12}$ for identical fermions.
The simultaneous symmetry eigenstates for bosons are now
\begin{equation*}
\Phi(\phi) \propto 
\begin{cases}
\cos{n\phi}, ~~~ n=0,2,4,\ldots, &\text{Even Parity}\\
\sin{n\phi}, ~~~ n=1,3,5,\ldots, &\text{Odd Parity} 
\end{cases}
,
\end{equation*}
and for fermions
\begin{equation*}
\Phi(\phi) \propto 
\begin{cases}
\sin{n\phi}, ~~~ n=2,4,6,\ldots, &\text{Even Parity} \\
\cos{n\phi}, ~~~ n=1,3,5,\ldots, &\text{Odd Parity}
\end{cases}
.
\end{equation*}
The range of integration can in this case be reduced by a factor of 4, and the
boundary conditions extracted as above.

\section{Summary of results for square-well and $\delta$-function interactions}
\label{app:Summary}

We summarize in this section the boundary and quantization conditions, along with
the large $R$ solutions, for all symmetries of three identical particles with
square-well interactions (see Table~\ref{table:eq_finite} and Sec.~\ref{SqWell}) and for all symmetries of
two identical particles with $\delta$-function interactions (see Table~\ref{table:uneq_mass} and Sec.~\ref{Two_delta}).

We expect the square-well results in Table~\ref{table:eq_finite} can be generalized
to an arbitrary short-range interactions in basically the same way as Eq.~(\ref{GenInt}).  That
is, the left-hand sides of the quantization conditions would get replaced by the 
two-body log-derivative.
\begin{table}
\caption{\label{table:eq_finite}
The boundary conditions, quantization conditions, and allowed asymptotic $\kappa$ 
for three identical particles with square-well interactions.  The
constants $\beta$ and $\phi_b$ are defined in Sec.~\ref{SqWell}.}
\begin{center}
\begin{tabular}{c}
\hline \hline
Even Parity Bosons  \\
$\Phi'(\frac{\pi}{2})=\Phi'(\frac{2\pi}{3})=0$ \\
$\sqrt{\kappa^2+\beta^2}\tan{\left[\sqrt{\kappa^2+\beta^2}(\phi_b - \frac{\pi}{2})\right]}=\kappa\tan{\left[\kappa(\phi_b-\frac{2\pi}{3})\right]} $ \\
$\kappa\underset{R\rightarrow\infty}{\longrightarrow}3,9,15...$\\
\hline
Odd Parity Bosons \\		  
$\Phi'(\frac{\pi}{2})=\Phi(\frac{2\pi}{3})=0$\\
$-\sqrt{\kappa^2+\beta^2}\tan{\left[\sqrt{\kappa^2+\beta^2}(\phi_b-\frac{\pi}{2})\right]}=\kappa\cot{\left[\kappa(\phi_b-\frac{2\pi}{3})\right]} $\\
$\kappa\underset{R\rightarrow\infty}{\longrightarrow}6,12,18...$\\
\hline
Even Parity Fermions\\
$\Phi(\frac{\pi}{2})=\Phi(\frac{2\pi}{3})=0$\\
$\sqrt{\kappa^2+\beta^2}\cot{\left[\sqrt{\kappa^2+\beta^2}(\phi_b-\frac{\pi}{2})\right]}=\kappa\cot{\left[\kappa(\phi_b-\frac{2\pi}{3})\right]} $\\
$\kappa\underset{R\rightarrow\infty}{\longrightarrow}6,12,18...$\\
\hline
Odd Parity Fermions\\
$\Phi(\frac{\pi}{2})=\Phi'(\frac{2\pi}{3})=0$\\
$\sqrt{\kappa^2+\beta^2}\cot{\left[\sqrt{\kappa^2+\beta^2}(\phi_b-\frac{\pi}{2})\right]}=-\kappa\tan{\left[\kappa(\phi_b-\frac{2\pi}{3})\right]} $\\
$\kappa\underset{R\rightarrow\infty}{\longrightarrow}3,9,15...$\\
\hline \hline
\end{tabular}
\end{center}
\end{table}

\begin{table*}
\caption{\label{table:uneq_mass}
The boundary conditions, quantization conditions and large $R$ solutions are shown for the two-identical-particle case.
The constants $\alpha_S$, $\alpha_D$, and $\phi_{23}$ are defined below Eq.~(\ref{uneq_vdelta}) and in 
Appendix~\ref{coordinates}.}
\begin{center}
\begin{tabular}{c}
\hline\hline
Even Parity Bosons  \\
$\Phi'(0)=0 \qquad\text{and}  \qquad \Phi'(\frac{\pi}{2}+\epsilon)=-\Phi'(\frac{\pi}{2}-\epsilon)$ \\
$\left(-\kappa + \frac{2}{\kappa}\mu^2 g^2 \alpha_S\alpha_D R^2\right)\tan\left[\kappa(\phi_{23}-\frac{\pi}{2})\right] - \mu g \alpha_S R \tan{\kappa\phi_{23}}\tan\left[\kappa(\phi_{23}-\frac{\pi}{2})\right] + \kappa\tan{\kappa\phi_{23}} = \mu g (\alpha_S + 2\alpha_D)R $\\ 
$\kappa=n\pi/|\phi_{23}-\frac{\pi}{2}|, ~ n=1,2,3...$  ~~~ and ~~~  $\kappa=(n+1/2)\pi/|\phi_{23}|,~ n=0,1,2...$\\
\hline
Odd Parity Bosons \\		  
$\Phi(0)=0 \qquad \text{and}  \qquad \Phi'(\frac{\pi}{2}+\epsilon)=-\Phi'(\frac{\pi}{2}-\epsilon)$ \\
$\left(-\kappa + \frac{2}{\kappa}\mu^2 g^2\alpha_S\alpha_D R^2\right)\tan\left[\kappa(\phi_{23}-\frac{\pi}{2})\right]
 - \mu g\alpha_S R \cot{\kappa\phi_{23}}
	\tan\left[\kappa(\phi_{23}-\frac{\pi}{2})\right] + \kappa\cot{\kappa\phi_{23}} = \mu g(\alpha_S + 2\alpha_D)R $\\
$\kappa=n\pi/|\phi_{23}-\frac{\pi}{2}|,~ n=1,2,3...$ ~~~  and ~~~  $\kappa=n\pi/|\phi_{23}|, ~ n=0,1,2...$\\
\hline
Even Parity Fermions\\
$\Phi(0)=0 \qquad  \text{and} \qquad \Phi(\frac{\pi}{2})=0$\\
$\kappa\cot\left[\kappa(\phi_{23}-\frac{\pi}{2})\right]-\kappa\cot{\kappa\phi_{23}}=2\mu g\alpha_D R$ \\
$\kappa=n\pi/|\phi_{23}-\frac{\pi}{2}|, ~ n=1,2,3...$ ~~~  and~~~   $\kappa=n\pi/|\phi_{23}|, ~ n=1,2,3...$\\
\hline
Odd Parity Fermions\\
$\Phi'(0)=0 \qquad \text{and} \qquad \Phi(\frac{\pi}{2})=0$ \\
$\kappa\cot\left[\kappa(\phi_{23}-\frac{\pi}{2})\right]+\kappa\tan{\kappa\phi_{23}}=2\mu g\alpha_D R$ \\
$\kappa=n\pi/|\phi_{23}-\frac{\pi}{2}|, ~ n=1,2,3...$  ~~~ and  ~~~ $\kappa=(n+\frac{1}{2})\pi/|\phi_{23}|, ~ n=0,1,2...$\\
\hline\hline
\end{tabular}
\end{center}
\end{table*}

\section{R-Matrix Propagation}
\label{app:rmatprop}

In order to solve for the multichannel $S$-matrix, we use the eigenchannel $R$-matrix
method coupled with the adiabatic hyperspherical representation.  The development closely
follows Refs.~\cite{mcrs} and \cite{burkejr1999tic,tolstikhin1998hec,baluja1982rmp}.

We begin with the variational expression 
\begin{equation}
\label{rrvar}
E=\frac{\int{d\phi RdR \;\Psi(R,\phi) H(R,\phi) \Psi(R,\phi)}} {\int{d\phi RdR \; |\Psi(R,\phi)|^2}},
\end{equation}
assuming $\Psi$ is real without loss of generality.
The Hamiltonian is given in Eqs.~(\ref{FullHam}) and (\ref{AdHam}).
Integrating the hyperradial kinetic energy by parts in Eq.~(\ref{rrvar}) gives
the variational expression,
\begin{equation}
\label{varb}
b = \frac{\int_{R_1}^{R_2}{d\phi R dR\left[-\diff{\Psi}{R}\diff{\Psi}{R} + \Psi (k^2 - 2\mu H_{\rm ad})\Psi\right)]}}
  {\int{d\phi \left[R_2 |\Psi(R_2,\phi)|^2 + R_1|\Psi(R_1,\phi)|^2\right]}},
\end{equation}
where $b$ is minus the log-derivative~\cite{greene,mcrs} normal to a hypersphere,
\begin{equation}
b=-\diff{\ln{\Psi}}{\hat{n}}.
\label{LogDeriv}
\end{equation}

In the adiabatic hyperspherical representation,
\begin{equation}
\label{channelexpansion}
\Psi(R,\phi)=\sum_iF_i(R)\Phi_i(R;\phi)
\end{equation}
where $\Phi_i(R;\phi)$ are eigenstates of Eq.~(\ref{AdHam}) with eigenvalues $U_i(R)$.  
Substitution of the expansion Eq.~(\ref{channelexpansion}) into Eq.~(\ref{varb}) results in a 
set of coupled equations for $F_i(R)$.  We further expand the radial functions in a basis set
as $F_i(R)=\sum_n{c_{ni} B_n(R)}$, converting the coupled equations
into the generalized eigenvalue problem
\begin{equation}
\label{geneig}
b\mathbf{\Lambda}\vec{c}=\mathbf{\Gamma}\vec{c}.
\end{equation}
The matrices $\mathbf{\Lambda}$ and $\mathbf{\Gamma}$ are defined as
\begin{widetext}
\begin{align}
\label{lambda_gamma}
\Lambda_{(m,i),(n,j)} &= \delta_{i,j}\!\sum_{n,m}{\left[R_1 B_n(R_1)B_m(R_1) + R_2 B_n(R_2)B_m(R_2)\right]} \\
\Gamma_{(m,i),(n,j)}&=\int_{R_{1}}^{R_{2}}dRR   \Biggl\{ -\delta_{i,j}\diff{B_m}{R}\diff{B_n}{R} - 
		P_{ij} \left(\diff{B_m}{R}B_n - B_m\diff{B_n}{R} \right)    
   + \mathbf{P}^2_{ij}B_m B_n +
		\delta_{i,j}B_m\left[k^2-2\mu U_i(R)\right]B_n  \Biggr \}  \nonumber.
\end{align}
\end{widetext}
We require the first-derivative nonadiabatic coupling,
\begin{equation}
\label{pmat}
P_{ij}(R)=\left<\!\!\left<\Phi_i(R;\theta)\left|\diff{}{R}\right|\Phi_j(R;\theta)\right>\!\!\right>,
\end{equation}
and the second derivative coupling,
\begin{equation}
\label{qmat}
Q_{ij}(R)=\left<\!\!\left<\Phi_i(R;\theta)\left|\secdiff{}{R}\right|\Phi_j(R;\theta)\right>\!\!\right>
\end{equation}
The double bracket notation implies integration only over $\phi$.  The square of $\mathbf{P}$ is given by the
symmetric part of $\mathbf{Q}$ through the relation $Q_{ij} = \mathbf{P}^2_{ij} + \diff{P_{ij}}{R}$:
\begin{equation}
\mathbf{P}^2_{ij}=-\left<\!\!\left<\diff{\Phi_i(R,\theta)}{R}\bigg| \diff{\Phi_j(R);\theta}{R}\right>\!\!\right>.
\end{equation}

Equation~(\ref{geneig}) is solved for the expansion coefficients $\vec{c}$, yielding a solution 
in the region $R_1 \le R \le R_2$.
For the first region $0 \le R\ \le R_1$, we require $F_i'(0) = 0$ but
impose no boundary condition at $R_1$.
The generalized eigenvalue problem then yields an eigenvalue $b_\beta$ and 
wave function $F_{i\beta}$ for each open or weakly closed channel at $R_1$.  

To propagate the solution to large $R$, we solve the system of equations again in the 
region $R_1 \le R \le R_2$ with no boundary 
conditions at either $R_1$ or $R_2$.  

A valid solution in the region $0 \le R \le R_2$ is then constructed by matching the
solutions from the two regions at $R_1$
and by requiring that the overall solution be an eigenchannel solution of the $R$-matrix,
\begin{equation}
\label{rmat}
\mathbf{R} = (\mathbf{F})(\mathbf{F'})^{-1}.
\end{equation}
That is, $\Psi$ should have constant normal derivative at the surface $R=R_2$.
This procedure is repeated until the solutions can be 
accurately matched to analytic asymptotic forms~\cite{baluja1982rmp,burkejr1999tic,tolstikhin1998hec}.  

To be more explicit, we write the full wave function
outside the $R$-matrix volume as
\begin{equation}
\Psi^{out}_{\beta}(R,\phi) \!= \!\sum_j\Phi_j(R;\phi)
\left[f_j(R)I_{j\beta} \!-\! g_j(R)J_{j\beta}\right],
\end{equation}
where $f_j(R)$ and $g_j(R)$ are 
\begin{align}
f_j(R) \!&= \!
\begin{cases}
\sqrt{\!\frac{2}{\pi k_{f}R}}\cos{k_{f}R}, &\text{if $j$ is a two-body channel}\\
J_\kappa(kR), &\text{if $j$ is a three-body channel}\\
\end{cases},\nonumber\\
g_j(R) \!&=\! 
\begin{cases}
\sqrt{\!\frac{2}{\pi k_{f}R}}\sin{k_{f}R}, &\text{if $j$ is a two-body channel}\\
Y_\kappa(kR), &\text{if $j$ is a three-body channel}\\
\end{cases}
.
\end{align}
The order $\kappa$ is determined as described in Sec.~\ref{section:threshold}.  
The solution inside the volume involves the numerical functions $F_{j\beta}$
\begin{equation}
\Psi^{in}_{\beta}(R;\phi) = \sum_j F_{j\beta}(R)\Phi_j(R;\phi).
\label{psi_in}
\end{equation}
The matrices $\mathbf{I}$ and $\mathbf{J}$ are obtained from matching at some
large distance $R=R_M$, which is conveniently accomplished using
\begin{align}
\label{mat_IJ}
I_{j\beta} &= W(g_j,F_{j\beta})/W(g_j,f_j) \nonumber\\
J_{j\beta} &= W(f_j,F_{j\beta})/W(g_j,f_j) 
\end{align}
where $W(f,g)$ denotes the Wronskian of $f$ and $g$.
Defining $(\mathbf{f})_{ij}=\delta_{ij}f_i$ and $(\mathbf{g})_{ij}=\delta_{ij}g_i$, the $K$-matrix is
\begin{equation}
\mathbf{K}(R_M) = (\mathbf{f}-\mathbf{f}'\mathbf{R})(\mathbf{g}-\mathbf{g}'\mathbf{R})^{-1}.
\end{equation}
The $R$-matrix is calculated via Eq.~(\ref{rmat}) using the numerical solutions $F_{i\beta}$
and the result $F_{i\beta}'(R_M) = -b_\beta F_{i\beta}(R_M)$.  The latter holds only
at large $R$ where $P_{ij}\rightarrow 0$ since the exact relation is
\begin{equation}
-b_\beta F_{i\beta} = F_{i\beta}' + \sum_j{P_{ij}F_{j\beta}}
\end{equation}
in which each quantity is evaluated at $R_M$.
This relation is obtained by differentiating Eq.~(\ref{psi_in}) and projecting
the result onto $\Phi_i$.
Finally, the $S$-matrix is calculated from $\mathbf{K}$ using
\begin{equation}
\mathbf{S}(R_M)=\frac{\mathbf{\openone}+i\mathbf{K}(R_M)}{\mathbf{\openone}-i\mathbf{K}(R_M)}.
\end{equation}

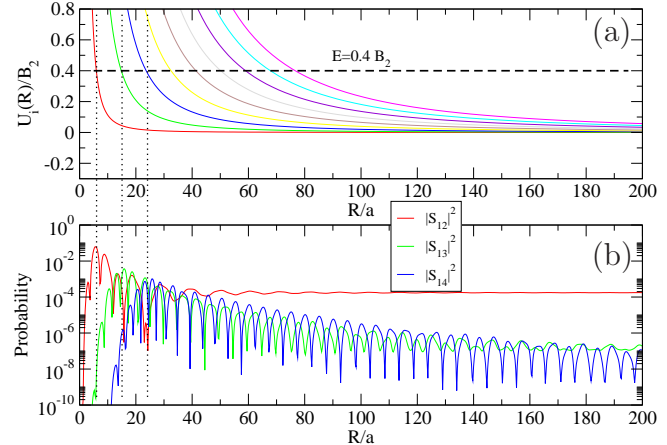
\begin{figure}[!t]
\begin{center}
\leavevmode
\includegraphics[width=3.375in,clip=true]{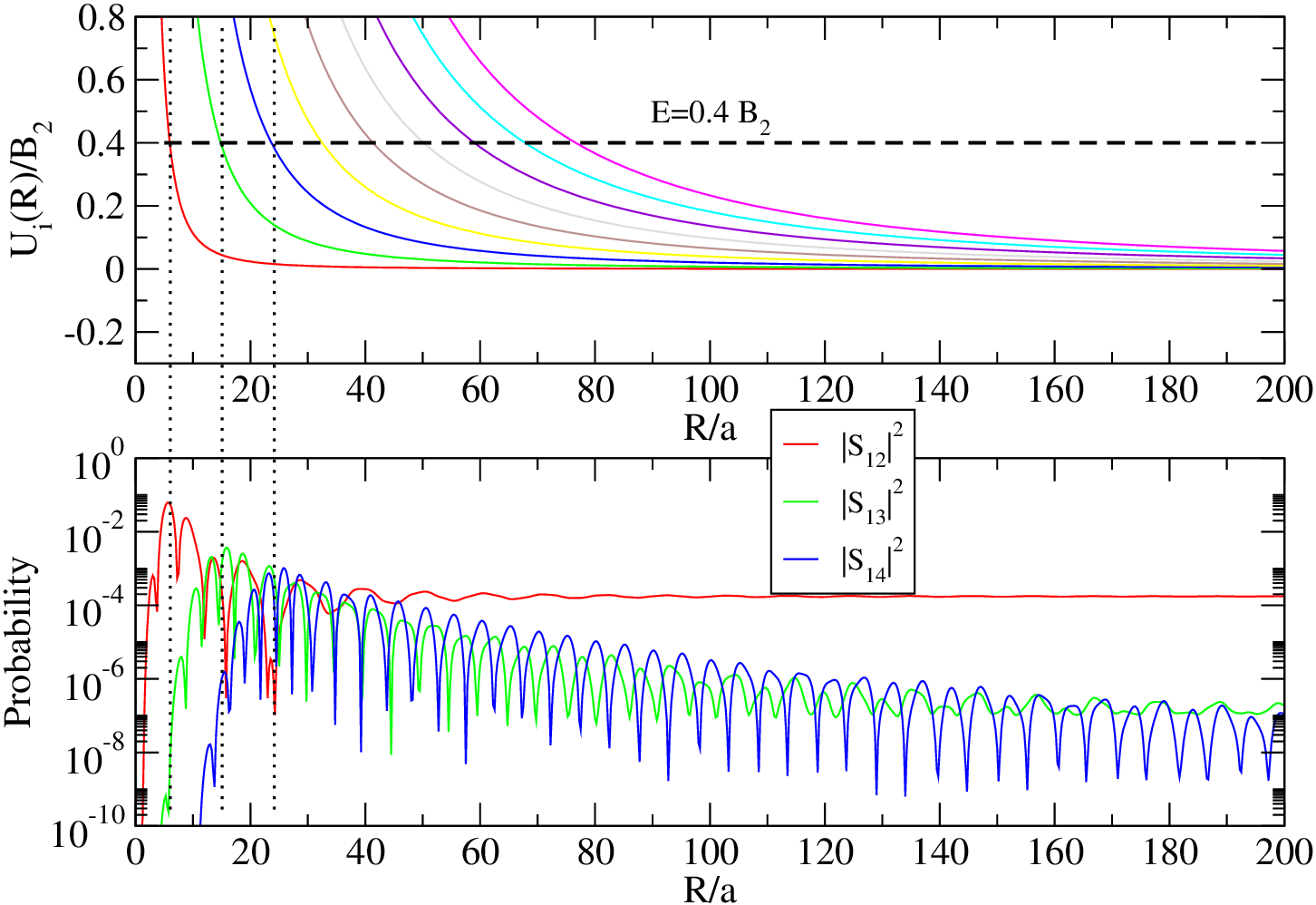}
 \begin{picture}(0,0)(0,0)
      \put(97,163){\large (a)}
      \put(97,78){\large (b)}
 \end{picture}
\caption{
(color online) (a) Three-body potential curves for the P\"oschl-Teller two-body potential with $L=2$ and $D=1/2$.
The horizontal dashed line indicates the collision energy, and the vertical dotted lines mark the classical turning point.
(b) The convergence of the transition probability between the two-body channel and the lowest three 
three-body channels at $E=0.4 B_2$ as a function of the matching distance.  
}
\label{fig:box_size}
\end{center}
\end{figure}

It is important to propagate the $R$-matrix to large $R$ in order to obtain a converged, 
unitary $S$-matrix.
In Fig.~\ref{fig:box_size}, we show the convergence of a few $S$-matrix elements with the
matching distance $R_M$ for an eight-channel 
calculation using the P\"oschl-Teller potential Eq.~(\ref{PT_pot}) with $L=2$ and
$D=1/2$.  Figure~\ref{fig:box_size}(a) shows the lowest nine 
potential curves corresponding to three-body channels.  (Since it converges to --1 on 
the scale of the figure, the two-body channel is not visible.)
The horizontal dashed line shows the energy at which the calculation in
Fig.~\ref{fig:box_size}(b) was carried out, and the vertical dotted lines mark the classical turning points for the first 
three channels.  Note that the probability $|S_{1i}(R_M)|^2$ peaks approximately at the classical turning point for $U_i(R)$.  

For the calculations presented in Fig.~\ref{fig:ZRP_vs_fin_elastic}, we propagated the $R$-matrix to $R=1000 a$ 
to assure convergence.  It is evident from Fig.~\ref{fig:box_size} that the quantity $|S_{12}|^2$ has already 
converged by $R=200 a$.  
We have also verified that this probability is stable with respect to the inclusion of more coupled channels.
The other probabilities, however, have not yet converged as well, although their magnitude
makes them negligible.

Our calculation of the matrix elements in Eq.~(\ref{lambda_gamma}) is simplified by using B-splines as our radial basis
set $\{B_n(R)\}$~\cite{deboor1978pgs}.  This choice also simplifies the imposition of boundary conditions since 
B-splines have only local support.  We typically use ten fifth-order 
B-splines within each $R$-matrix sector, leading to a $(10 \times N)\times(10 \times N)$ matrix equation 
($N$ is the number of channels).
The size of the sectors $R_{i+1}-R_i$ is chosen to be no more than one de Broglie wavelength in the lowest (two-body) channel.

\bibliography{3brecomb}

\end{document}